\documentclass[a4paper,12pt]{article}
\usepackage[utf8x]{inputenc}
\usepackage{graphicx} 
\usepackage{epsfig}
\usepackage{amssymb}
\usepackage{amsmath}
\usepackage{amsfonts}
\usepackage{mathrsfs}
\usepackage[english]{babel}   

\usepackage{fancyhdr}
\usepackage{lastpage}
\usepackage{a4wide}
\usepackage[normal]{subfigure}
\usepackage{color}
\usepackage{inputenc}
\usepackage{hyperref}

 \usepackage{empheq}

\usepackage[section]{placeins}
 
\newtheorem{theo}{Theorem}[section]

\newtheorem{comentario}{{\bf Note}}[section]

\def\d{\displaystyle}

\def\mo{\mathcal{O}}
\def\ep{\varepsilon}

\def\pt{\partial_t}
\def\px{\partial_x}

\def\bt{\tilde{b}}
\def\st{\boldsymbol{U}}
\def\sst{\boldsymbol{U}}
\def\tnh{\boldsymbol{\mathcal{T}}_{NH}}
\def\pnh{\tilde {p_1}_{|b+h_2}}
\def\bnh{\boldsymbol{\mathcal{B}}}

\def\sg{\textrm{sgn}}

\def\div{\textrm{div}}
\def\eff{\textrm{eff}}
\def\sgn{\textrm{sgn}}
\def\df{\vartheta}

\title{A two-layer shallow water model for bedload sediment transport: convergence to Saint-Venant-Exner model}

\author{C. Escalante \thanks{Dpto. de A.M., E. e I.O., y Matem\'atica Aplicada - Universidad de M\'alaga, Campus de Teatinos s/n. 29071 M\'alaga, Spain. {\tt  (escalante@uma.es)}}, E.D. Fern\'andez-Nieto \thanks{Dpto. Matem\'atica Aplicada I. ETS Arquitectura - Universidad de Sevilla.
  Avda. Reina Mercedes N. 2. 41012-Sevilla, Spain. {\tt (edofer@us.es), (gnarbona@us.es)}}, \\\ T. Morales de Luna \thanks{Dpto. de Matem\'aticas. Universidad de C\'ordoba. Campus de Rabanales. 14071 C\'ordoba, Spain. {\tt (tomas.morales@uco.es)}}, 
  G. Narbona-Reina\footnotemark[2]
}

\begin{document}
\maketitle

\begin{abstract}

A two-layer shallow water type model is proposed to describe bedload sediment transport. The upper layer is filled by water and the lower one by sediment. The key point falls on the definition of the friction laws between the two layers, which are a generalization of those introduced in Fern\'andez-Nieto et al. ({\it ESAIM: M2AN}, 51:115-145, 2017). This definition allows to apply properly the two-layer shallow water model for the case of intense and slow bedload sediment transport. Moreover, we prove that the two-layer model converges to a Saint-Venant-Exner system (SVE) including gravitational effects when the ratio between the hydrodynamic and morphodynamic time scales is small.  The SVE with gravitational effects is a degenerated nonlinear parabolic system. This means that its numerical approximation is very expensive from a computational point of view, see for example T. Morales de Luna et al.  ({\it J. Sci. Comp.}, 48(1): 258--273, 2011). In this work, gravitational effects are introduced into the two-layer system without such extra computational cost. Finally, we also consider a generalization of the model that includes a non-hydrostatic pressure correction for the fluid layer and the boundary condition at the sediment surface. Numerical tests  show that the model provides promising results and behave well in low transport rate regimes as well as in many other situations. 

\end{abstract}

\newpage

\section{Introduction}

One of the difficulties to simulate bedload
sediment transport is that the interaction between fluid and sediment is
very small. Bedload is the type of transport where sediment grains roll  along the bed or single grains jump over the bed a length proportional to their diameter, losing for instants the contact with the soil. This is opposed to transport in suspension where particles are transported by the fluid without touching the bottom for long period of times. For low Froude numbers, bedload is the dominating transport mechanism. In this case the characteristic velocities of fluid and sediment are very
different. The bedload transport rate is low and the characteristic velocity of the sediment is much smaller than that of the fluid. 

In \cite{FerMoNarZab}, a multi-scale analysis is performed taking into account that the velocity of the sediment layer is smaller than the one of the fluid layer. This leads to a shallow water type system for the fluid layer and a lubrication Reynolds equation for the sediment one.  

For the case of uniform flows the thickness of the moving sediment layer can be predicted, because erosion and deposition rates are equal in those situations. This is a general hypothesis that is assumed when modeling bedload transport. The usual approach is to consider a coupled system consisting of a Shallow Water system for the hydrodynamical part combined with a morphodynamical part given by the so-called Exner equation. The whole system is known as Saint Venant Exner (SVE) system \cite{exner}. Exner equation depends on the definition of the solid transport discharge. Different classical definitions can be found for the solid transport discharge, for instance the ones given by  Meyer-Peter $\&$ M\"uller \cite{MeyPetMul}, Van Rijn's \cite{Van},  Einstein \cite{Ein}, Nielsen \cite{Nie}, Fern\'andez-Luque $\&$ Van Beek \cite{FerLuq}, Ashida $\&$ Michiue \cite{ashida_michiue}, Engelund $\&$ Fredsoe \cite{engelund_fredsoe}, Kalinske \cite{Kal}, Charru \cite{Charru}, etc. A generalization of these classical models was introduced in \cite{FerMoNarZab} where the morphodynamical component is deduced from a  Reynolds equation and includes gravitational effects in the sediment layer. Classical models do not take into account in general such gravitational effects because in their derivation the hypothesis of nearly horizontal sediment bed is used (see for example \cite{kovacs_parker}).\\
In general, classical definitions for solid transport discharge can be written as follows,
\begin{equation} \label{eq_qb_gen}
\frac{q_b}{Q}= \mbox{sgn}(\tau) \,\frac{ k_1}{(1-\varphi)} \,  \theta^{\, m_1} \, (\theta-k_2 \, \theta_c)_+^{m_2} \left( \sqrt{\theta}- k_3 \, \sqrt{\theta_c} \right)^{m_3}_+,
\end{equation}
where $Q$ represents the characteristic discharge, $Q=d_s \sqrt{g (1/r-1)d_s}$,  $r=\rho_1/\rho_2$ is the density ratio,  $\rho_1$ being the fluid density and $\rho_2$ the density of the sediment particles; $d_s$ the mean diameter of the sediment particles, and $\varphi$ is the averaged porosity. 
The coefficients $k_l$ and $m_l$, $l=1,2,3$, are positive constants that depend on the model. We usually find $m_2=0$ or $m_3=0$, for example, Meyer-Peter $\&$ M\"uller model takes $m_3=0$ and Ashida $\&$ Michiue's model uses $m_2=0$.\\
The Shields stress, $\theta$, is defined as the ratio between the agitating and the stabilizing forces, $\theta= |\tau| d_s^2 / ( g (\rho_2 - \rho_1) d_s^3)$, $\tau$ being the shear stress at the bottom.  For example, for Manning's law, we have $\tau=\rho_1 g h_1 n^2 u_1 |u_1| / h_1^{4/3}$. Where $h_1$ and $u_1$ are the thickness and the velocity of the fluid layer, respectively, and $n$ is the Manning coefficient.\\
Finally, $\theta_c$ is the critical Shields stress. The positive part, $( \, \cdot \, )_+$, in the definition implies that the solid transport discharge is not null only if  $\theta > k \theta_c$ (with $k=k_2$ when $m_2>0$ and $k=\sqrt{k_3}$ when $m_3>0$).  If the velocity of the fluid is zero, $u_1=0$, we have $\theta =0 < k \theta_c$, and for any model that can be written under the structure (\ref{eq_qb_gen}) we obtain that $q_b=0$, which means that there is no movement of the sediment layer. This is even true when the sediment layer interface is not horizontal which is a consequence of the fact that classical models do not take into account gravitational effects.

In order to introduce gravitational effects in classical models, Fowler et al. proposed in \cite{FowKopOak} a modification of the Meyer-Peter \& M\"uller formula that consist  in replacing $\theta$ by $\theta_{\eff}$, where:
\begin{equation} \label{eq_theta_eff_fowler}
\theta_{\eff}= \left| \sgn(u_1) \theta - \df \partial_x (b+h_2) \right|,
\end{equation}
with
\begin{equation} \label{def_vartheta}
 \df=\frac{\theta_c}{\tan \delta } ,
\end{equation}
$\delta$ being the angle of repose of the sediment particles. The sediment surface is defined by $z=b+h_2$, where $h_2$ is the thickness of the sediment layer and $b$ the topography function or bedrock layer. Then, $\theta_{\eff}$ is defined in terms of the gradient of sediment surface. This is a definition that can be also considered for 2D simulations, because in this case  $\theta_{\eff}$ is defined as the norm of the vector $\theta u_1/\|u_1\| - \df \nabla (b+h_2)$. Other alternatives have been proposed in the literature, namely consisting of a modification of  $\theta_c$, instead of a modification of $\theta$. Both approximations are equivalent only in some cases for 1D problems (see \cite{FerMoNarZab}). Moreover, defining the modification of $\theta_c$ for arbitrarily sloping bed for 2D problems is not an easy task (see \cite{parker_2003}, \cite{seminara_2002}).

As we mentioned previously, in \cite{FerMoNarZab} a SVE model is deduced from a multiscale analysis. The model includes gravitational effects and the authors deduce that it can also be seen as a modification of classical models by replacing $\theta$ by the proposed values $\theta^{(L)}_{\eff}$ or $\theta^{(Q)}_{\eff}$ depending if we set a linear or a quadratic friction law respectively between the fluid and the sediment layer.  For the case of nearly uniform flows, the thickness of the sediment layer which corresponds to moving particles, $h_m$, is of order of $d_s/\df$. Under this assumption, for the case of a linear friction law, the definition of the effective shear stress proposed in \cite{FerMoNarZab} can be written as follows:

\begin{equation} \label{eq_theta_eff_L}
\theta^{(L)}_{\eff}= \left| \sgn(u_1)\theta- \df   \partial_x (b+h_2)- \df \frac{ \rho_1 }{\rho_2-\rho_1}  \partial_x(b+h_1+h_2)\right|.
\end{equation}
Let us remark that if the water free surface is horizontal, the definition of $\theta^{(L)}_{\eff}$ coincides with $\theta_{\eff}$ (\ref{eq_theta_eff_fowler}), proposed by Fowler et al. in \cite{FowKopOak}. Otherwise, the main difference is that this definition for the effective shear stress takes into account not only the gradient of the sediment surface but also the gradient of the water free surface.

For the case of a quadratic friction law, although the definition is a combination of the same components, it is rather different. In this case we can write the effective Shields parameter proposed in \cite{FerMoNarZab} as follows:
\begin{equation} \label{eq_theta_eff_Q}
\theta_{\eff}^{(Q)} = \left| \sg(u_1)   \sqrt{\theta}-\sqrt{\frac{\df \rho_1}{\rho_2-\rho_1} |\partial_x(\frac{\rho_1}{\rho_2} h_1+ h_2+b) |} \, \, \sg\left(\partial_x(\frac{\rho_1}{\rho_2} h_1+ h_2+b)\right) \right|^2.
\end{equation}
In the case of submerged bedload sediment transport, the drag term is defined by a quadratic friction law. Thus, we should consider an effective Shields stress given by $\theta_{\eff}^{(Q)}$.  Nevertheless, in the bibliography it is $\theta_{\eff}$ (\ref{eq_theta_eff_fowler}) which is usually considered, regardless the fact that $\theta_{\eff}$ is an approximation of $\theta_{\eff}^{(L)}$ which is deduced from a linear friction law. 

Although the quantities involved in the definitions of the effective Shields parameter associated to linear or quadratic friction are the same, their values may be very different. They verify
$$
|\theta_{\eff}^{(L)}-\theta_{\eff}^{(Q)}| =  {\mathcal O}\left( \, | \partial_x(\frac{\rho_1}{\rho_2} h_1+ h_2+b) |  \left( \sqrt{\theta}-\sqrt{\frac{\df \rho_1}{\rho_2-\rho_1} |\partial_x(\frac{\rho_1}{\rho_2} h_1+ h_2+b) |} \,\, \right) \, \right).
$$
For instance, if we consider a initial condition with water at rest and a high gradient in the sediment surface, the difference between $\theta_{\eff}^{(L)}$ and $\theta_{\eff}^{(Q)}$ is of order of the gradient of the sediment surface. Thus, in the framework of SVE model, the definition  $\theta_{	\eff}^{(Q)}$ should be considered in order to be consistent with the quadratic friction law  usually considered for the drag force between the fluid and the sediment.
\bigskip

In any case, considering the definitions  $\theta_{\eff}$ (\ref{eq_theta_eff_fowler}), $\theta_{\eff}^{(L)}$ (\ref{eq_theta_eff_L}), or $\theta_{\eff}^{(Q)}$ (\ref{eq_theta_eff_Q}),  means that the corresponding SVE system with gravitational effects is a parabolic degenerated partial differential  system with non linear diffusion. Moreover, the system cannot be written as combination of a hyperbolic part plus a diffusion term. 

Let us remark that in the literature a linearized version  can be found, where gravitational effects are included by considering a classical SVE model with an additional viscous term, see for example \cite{TasRheVio} and references therein. The drawback of this approach is that the diffusive term should not be present in stationary situations, for instance when the velocity is not high enough and sediment slopes are under the one given by the repose angle. In such situations it is necessary to include some external criteria that controls whether the diffusion term is applied or not. This is not the case in definitions   (\ref{eq_theta_eff_L}) or (\ref{eq_theta_eff_Q})    where the effective Shields stress is automatically limited by the effect of the Coulomb friction angle. 

From the computational point of view, any of previous strategies for gravitational effects are expensive. Due to the parabolic nature of the equation, explicit methods have a very restrictive CFL condition, and implicit methods need to incorporate a fixed point algorithm for its resolution (cf. \cite{morales2011duality}). 

\bigskip

Another approach used to study bedload sediment transport is to consider two-layer shallow water type model, see for example \cite{spinewine_tesis}, \cite{swart}, \cite{savary_tesis}.  Nevertheless they are usually proposed in the case of high bedload transport rate, such as in dam break situations. This differs from classical SVE system where low transport rates with small Froude number is assumed. 

In this paper we propose a two-layer shallow water model that  behaves as a SVE model when the bedload regime holds. It has the advantage that it converges to a generalization of  SVE model with gravitational effects for low transport regimes while being valid for higher transport regimes as well. Moreover, it has the advantage that the inclusion of gravitational effects does not imply any extra computational effort, opposed to what happens for classical SVE systems with gravity effects.

\medskip

An additional advantage of the model introduced here is that it will take into account dispersive effects. When modelling and simulating geophysical shallow flows, the nonlinear shallow water equations are often a good choice as an approximation of the Navier-Stokes equations. Nevertheless, they are derived by assuming hydrostatic pressure and they do not take into account non-hydrostatic effects or dispersive waves. In coastal areas, close to the continental shelf, non-hydrostatic effects or equivalently, dispersive waves may become important.

 In recent years, effort has been done in the derivation of relatively simple mathematical models for shallow water flows that include long nonlinear water waves. See for instance the works in~\cite{jacques,madsen,peregrine,yamazaki} among others. The hypothesis is that this dispersive effects will have an important impact on the sediment layer.

Following this idea, in this work we will consider a non-hydrostatic pressure for the fluid layer.  The non-hydrostatic pressure act on the sediment layer as a boundary condition on the interface between the fluid and the sediment. As it will be shown, these non-hydrostatic effects have an important impact when compared with the hydrostatic version.
\bigskip

The paper is organized as follows: The two-layer Shallow Water model that we propose is introduced in Section 2. The energy balanced verified by the proposed model is also stated. Section 3 is devoted to show the convergence of the model to the SVE model with gravitational effect proposed in \cite{FerMoNarZab}. In Section 4 we present the generalization of the proposed model by including non-hydrostatic pressure in the fluid layer and its influence on the sediment layer as a modification of the gradient pressure at the interface. A numerical method to approximate this model is described in Section 5. Three numerical tests are shown in Section 6. Finally, conclusions are presented in Section 7.

\section{Proposed model}

We consider a domain with two immiscible layers corresponding to water (upper layer) and sediment (lower layer). The sediment layer is in turn decomposed into a moving layer of thickness $h_m$ and a sediment layer that does not move of thickness $h_f$, adjacent to the fixed bottom. These thicknesses are not fixed because there is an exchange of sediment material between the layers. Particles are eroded from the lower sediment layer and come into motion in the upper sediment layer. Conversely, particles from the upper layer are deposited into the lower sediment layer and stop moving.

\begin{figure}
\begin{center}
\includegraphics[width=10.cm]{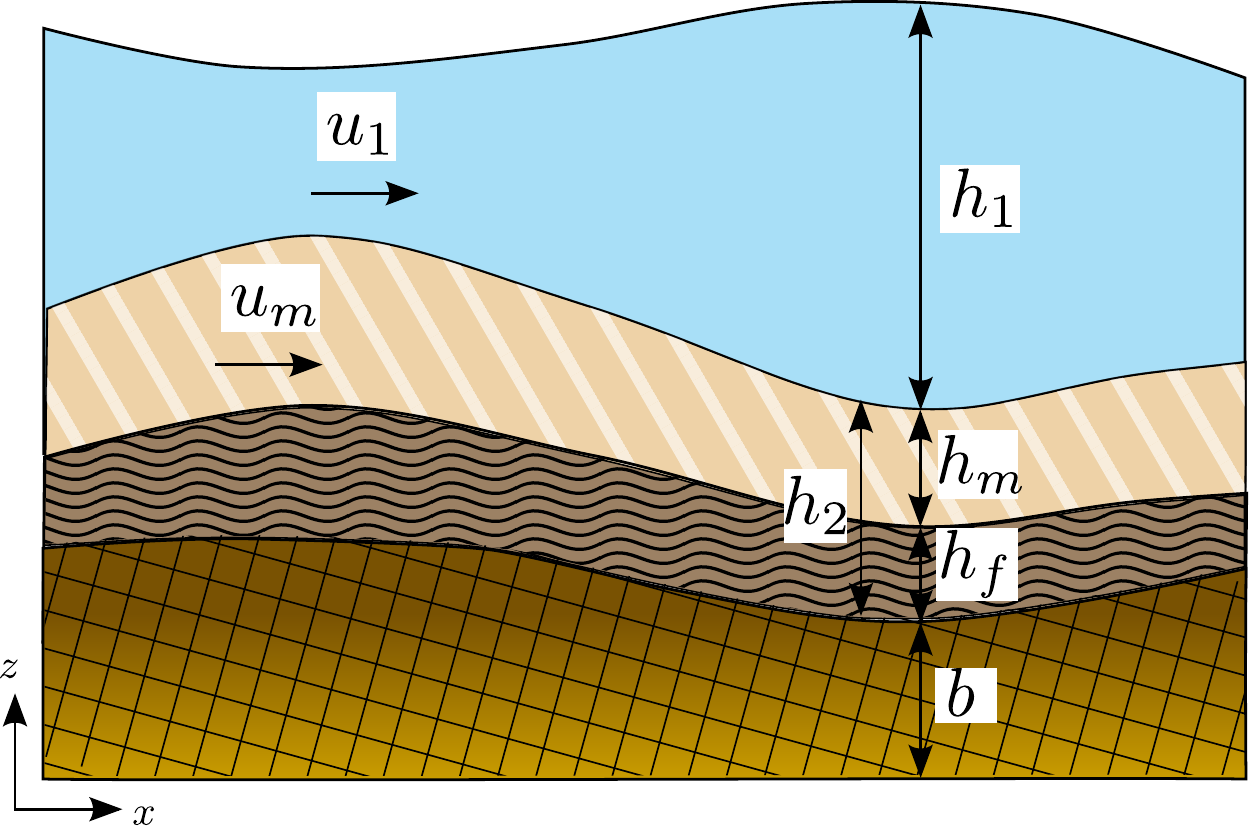}
\end{center}
\caption{Sketch of the domain for the fluid-sediment problem}
\label{fig:domain}
\end{figure}

We propose a 2D shallow water model that may be obtained by averaging on the vertical direction the Navier-Stokes equations and taking into account suitable boundary conditions. In particular, at the free surface we impose kinematic boundary conditions and vanishing pressure; at the bottom a Coulomb friction law is considered. The friction between water and sediment is introduced through the term $F$ at the water/sediment interface and the mass transference term in the internal sediment interface is denoted by $T$.
The general notation for the water layer corresponds to the subindex 1 and for the sediment layer to the subindex 2. 
Thus, the water of layer has a thickness $h_1$ and moves with horizontal velocity $u_1$. The thickness of the total sediment layer is denoted by $h_2=h_f+h_m$, and the moving sediment layer $h_m$ flow with velocity $u_m$. The fixed bottom is denoted by $b$.
See Figure \ref{fig:domain} for a sketch of the domain.\\
Note that the velocity of the sediment layer is defined as $u_2=u_m$ in the moving layer and $u_2=0$ in the static layer. We assume first an hydrostatic pressure regime.
\\\\
Then we propose the following two-layer shallow water model:
\begin{subequations}\label{bilayer_model}
\begin{equation}\label{mass1}
\pt h_1 +\div (h_1 u_1)=0
\end{equation}
\begin{equation}\label{mom1}
\pt(h_1 u_1) +\div(h_1 u_1\otimes u_1)+g h_1 \nabla_x(b+h_1+h_2)=-F
\end{equation}
\begin{equation}\label{mass2}
\pt h_2 +\div (h_m u_m)=0
\end{equation}
\begin{equation}\label{mom2}
\pt(h_m u_m) +\div(h_m u_m\otimes u_m)+g h_m \nabla_x(b+rh_1+h_2)=r F+\frac 1 2 u_m T-(1-r) gh_m \sg(u_m)\tan\delta
\end{equation}
\begin{equation}\label{mass3}
\pt h_f=-T
\end{equation}
\end{subequations}
where $r=\rho_1/\rho_2$ is the ratio between the densities of the water, $\rho_1$, and the sediment particles, $\rho_2$. $\delta$ is the internal Coulomb friction angle. In the next lines we give the closures for the friction term $F$ and the mass transference $T$.\\\\
Following \cite{FerMoNarZab} we consider two types of friction laws: linear and quadratic. The friction term for the linear friction law is defined as
\begin{equation}\label{fric_l}
F_L=C_L(u_1-u_m)\quad \textrm{with} \quad C_L=g\Big(\frac 1 r -1\Big)\frac{h_1 h_m}{\vartheta (h_1+h_m)\sqrt{(\frac 1 r -1)g d_s}}
\end{equation}
and for the quadratic friction law,
\begin{equation}\label{fric_q}
F_Q=C_Q(u_1-u_m)|u_1-u_m|\quad \textrm{with} \quad C_Q=\frac 1 \alpha \, \frac{h_1 h_m}{\vartheta (h_1+h_m)}.
\end{equation}
$d_s$ being the mean diameter of the sediment particles. $\vartheta$ is defined by equation (\ref{def_vartheta}). This definition of $\vartheta$ verifies the analysis of Seminara et al. \cite{seminara_2002}, who concluded that the drag coefficient is proportional to $\tan(\delta)/\theta_c$.

Remark that the calibration coefficient $\alpha$ has units of length so that $C_Q$ is non-dimensional. In \cite{FerMoNarZab}, $\alpha=d_s$ was assumed for the bedload in low transport situations. In our case, given that we deal with a complete bilayer system, this value is not always valid.
In bedload framework, we can establish from experimental observations that the region of particles moving at this level is at most 10-20 particle-diameter in height \cite{hubert}.\\ 
So we may assume that the thickness of the bed load layer is  $h_m=k \, d_s$ with $k\in[0,k_{\max}]$ ($k_{\max}=10$ or 20). So that, when $h_m\leq k_{\max} d_s$ we are in a bedload low rate regime and it makes sense to consider the friction coefficient as in \cite{FerMoNarZab}, that is, of the order of $d_s$. Conversely, when $h_m>k_{\max} d_s$ we are not in a bedload regime and then we must turn to a more appropriate friction coefficient. Thus, to be consistent with our previous work, we propose to take:
$$
\alpha=\left\{
\begin{array}{lc}
h_m & \textrm{if } h_m>k_{\max} d_s \\
 d_s \quad & \textrm{if } h_m\leq k_{\max} d_s
\end{array}
\right.
$$
Another possibility for the second case, would be to define $\alpha=k_{\max} d_s$ when $h_m\leq k_{\max} d_s$.  The coefficient $k_{\max}$ can be then considered as a calibration constant for the friction law.\\

\bigskip

The mass transference between the moving and the static sediment layers $T$  is defined in terms of the difference between the erosion rate, $\dot{z}_e$, and  the deposition rate $\dot{z}_d$. There exists in the literature different forms to close the definition of the erosion and deposition rates (see for example \cite{Charru}).  We consider in this work the following definition (see \cite{FerLuMoCor}):

$$
T =  \dot{z}_e - \dot{z}_d  \quad \mbox{with} \quad
\dot{z}_e = K_e (\theta_e - \theta_{c})_+\displaystyle\frac{\sqrt{g(1/r-1)d_s}}{1-\varphi}, \quad \dot{z}_d = K_d h_m \displaystyle\frac{\sqrt{g(1/r-1)d_s}}{d_s}.
$$
The coefficients $K_e$ and $K_d$ are erosion and deposition constants, respectively, $\varphi$ is the porosity. For the case of nearly flat sediment bed it is usually set $\theta_e=\theta$, which corresponds to the Bagnold's relation (see \cite{bagnold1956flow}). Nevertheless, to take into account the gradient of the sediment bed $\theta_e$ must be defined in terms of the effective Shields stress (see \cite{FerMoNarZab}). Then we define $\theta_e$ in terms of the friction law between the fluid and the sediment layers:
$$
\theta_e= \left\{ \begin{array}{l}
\theta_{\eff}^{(L)}, \mbox{defined in equation (\ref{eq_theta_eff_L}), for a linear friction law},\\ \\
\theta_{\eff}^{(Q)}, \mbox{defined in equation (\ref{eq_theta_eff_Q}), for a quadratic friction law}.
\end{array} \right.
$$

\bigskip

The proposed model has an exact dissipative energy balance, which is an easy consequence of two-layer shallow water systems, opposed to classical SVE models which have not. We obtain the following result.
\begin{theo}
System \eqref{bilayer_model} admits a dissipative energy balance that reads:
\begin{eqnarray}
\partial_t \left(rh_1 \frac{|u_1|^2}{2} + h_m \frac{|u_m|^2}{2} + \frac12 g (rh_1^2+h_2^2) +g\,rh_1 h_2 + gb(rh_1+h_2)\right) \nonumber
\\
+\div\left( r h_1 u_1 \frac{|u_1|^2}{2} + h_m u_m \frac{|u_m|^2}{2} +g\,rh_1 u_1 (h_1+h_2+b) +g h_m u_m(rh_1+h_2+b) \right)\nonumber \\
\leq -r (u_1-u_m) F-(1-r) g h_m|u_m| \tan\delta;\nonumber
\end{eqnarray}
where the friction term $F$ is given by \eqref{fric_l} or \eqref{fric_q}.
\end{theo}

The proof of the previous result is straightforward and for the sake of brevity we omit it. Notice that classical SVE model does not verify in general a dissipative energy balance. In \cite{FerMoNarZab} a modification of a class of classical SVE models and a generalization, by including gravitational effects, has been proposed that allows them to verify a dissipative energy balance. In the following section we see that the proposed two-layer model converges to the SVE model proposed in \cite{FerMoNarZab}.

\section{Convergence to the classical SVE system for small morphodynamic time scale}\label{sec:sve}
In this section we prove the convergence of system (\ref{bilayer_model}) to the Saint-Venant-Exner model presented in \cite{FerMoNarZab}. This model is obtained from an asymptotic approximation of the Navier-Stokes equations. In particular, it has the following advantages: 
\begin{itemize}
  \item it preserves the mass conservation,
  \item the velocity (and hence, the discharge) of the bedload layer is explicitly deduced,
  \item it has a dissipative energy balance.
\end{itemize}
The model introduced in \cite{FerMoNarZab} reads as follows:
\begin{equation}\label{svemodel}
\left\{\begin{array}{l}
\d \partial_t {h_1}  + \div_x q_1 = 0,\\\\
\d\partial_t q_1+ \div_x(h_1(u_1\otimes u_1) )+gh_1\nabla_x(b+h_2+h_1) = -\frac{gh_m}{r} \mathcal{P},\\\\
\d \partial_t {h_2} + \div_x\left( {h_m}\, v_b\,  \sqrt{(1/r-1) g d_s} \right)=0,\\\\
\partial_t h_f=-T_m.
\end{array}\right.
\end{equation}
with
\begin{equation}\label{defP1}
\mathcal{P}=\nabla_x(r h_1+ h_2+b)+(1-r)\sgn(u_2) \tan\delta .
\end{equation}
The definition of the non-dimensional sediment velocity $v_b$ depends on the friction law. When a linear friction law is considered, it reads:
\begin{equation}\label{vblf}
v_b^{(LF)}=\frac{u_1}{\sqrt{(1/r-1) g d_s}} 
- \frac{\df}{1-r} \mathcal{P},
\end{equation}
where 
$$
\sgn(u_2)= \displaystyle \sg \left( \frac{u_1}{\sqrt{(1/r-1) g d_s}} 
- \frac{\df}{1-r} \nabla_x(r h_1+ h_2+b) \right).
$$
For a quadratic friction law:
\begin{equation}\label{vbqf}
v_b^{(QF)}=\frac{u_1}{\sqrt{(1/r-1) g d_s}} 
- \Big(\frac{\df}{1-r}\Big)^{1/2} \, |\mathcal{P}|^{1/2} \sg(\mathcal{P}),
\end{equation}
where $\sgn(u_2)=\sgn(\Psi)$ and
$$
\Psi=\frac{u_1}{\sqrt{(1/r-1) g d_s}} -\left|\frac{\df}{1-r} \nabla_x(r h_1+ h_2+b)\right|^{1/2}\sg\left(\frac{\df}{1-r} \nabla_x(r h_1+ h_2+b)\right).
$$
\medskip

The convergence is obtained when we assume the adequate asymptotic regime in terms of the time scales. As it is well known, for the bedload transport problem, the morphodynamic time is much smaller than the hydrodynamic time, which 
makes the pressure effects much more important than the convective ones. As a consequence, the behavior of the sediment layer is just defined by the solid mass equation (Exner equation), omitting a momentum equation. This small morphodynamic time turn into an assumption of a smaller velocity for the lower layer. In order to fall into the bedload transport regime we must also assume that the thickness of the bottom layer is smaller, because it represents the layer of moving sediment. Thus, we suppose:
\begin{equation}\label{asympt}
u_m=\ep_u \tilde u_m;\quad h_m=\ep_h \tilde h_m; \quad T=\ep_u \tilde T.
\end{equation}
Now we take these values into the momentum conservation equation for the lower layer in  (\ref{bilayer_model}):
$$
\begin{array}{l}
\d \pt(\ep_h\ep_u \tilde h_m \tilde u_m) +\div(\ep_h\ep_u^2 \tilde h_m \tilde u_m\otimes \tilde u_m)+g \ep_h \tilde h_m \nabla_x(b+rh_1+h_2)\\
\hfill
\d = r \tilde F+\ep_u^2\frac 1 2 \tilde u_m \tilde T-(1-r) g\ep_h \tilde h_m \sg(\tilde u_m)\tan\delta
\end{array}
$$
Then, if we neglect second order terms, we get 
$$
g \ep_h \tilde h_m \nabla_x(b+rh_1+h_2)= r \tilde F-(1-r) g\ep_h \tilde h_m \sg(\tilde u_m)\tan\delta.
$$
In dimension variables, this equation reads:
$$
r  F=g  h_m \nabla_x(b+rh_1+h_2)+(1-r) g h_m \sg( u_m)\tan\delta=g h_m\mathcal{P}, 
$$
where the last equality follows from the definition of $\mathcal{P}$.
Thus the expression of the friction term is
\begin{equation}\label{frictionP}
r F={g h_m}\mathcal{P};
\end{equation}
which coincides with the friction term in the momentum equation of layer 1, r.h.s. of (\ref{svemodel}).\\
Now, from this equation and using the expressions of $F$, for linear \eqref{fric_l} and quadratic \eqref{fric_q} laws, we have to compute the value of $u_m$ to check that it fits with \eqref{vblf} and \eqref{vbqf} respectively.\\\\
$\circ$ Linear friction law:
\begin{eqnarray}
\tilde F&=&g\Big(\frac 1 r -1\Big)\frac{\ep_h h_1 \tilde h_m}{\vartheta (h_1+\ep_h \tilde h_m)\sqrt{(\frac 1 r -1)g d_s}} (u_1-\ep_u \tilde u_m) \nonumber \\
&=&g\Big(\frac 1 r -1\Big)\frac{1}{\vartheta \sqrt{(\frac 1 r -1)g d_s}}\, \frac{\ep_h \tilde h_m}{1+\ep_h \frac{\tilde h_m}{ h_1}} (u_1-\ep_u \tilde u_m) \nonumber \\
&=& g\Big(\frac 1 r -1\Big)\frac{\ep_h \tilde h_m}{\df \sqrt{(\frac 1 r -1)g d_s}}(u_1-\ep_u \tilde u_m)+\mo(\ep_h^2)
\end{eqnarray}
where in the last equality we have used that $\d \frac{1}{1+\ep_h \frac{\tilde h_m}{ h_1}}=1-\ep_h \frac{\tilde h_m}{ h_1}+\mo(\ep_h^2)$.\\
So turning to the dimension variables and neglecting second order terms, the equation \eqref{frictionP} reads:
$$
r  g\Big(\frac 1 r -1\Big)\frac{h_m}{\df \sqrt{(\frac 1 r -1)g d_s}}(u_1- u_m)=gh_m\mathcal{P}.
$$
From where we directly obtain that $u_m=v_b^{(LF)}\sqrt{(\frac 1 r -1)g d_s}$.\\\\
$\circ$ Quadratic friction law:\\
Note that in this case $\alpha$ reduces to $d_s$ and then
\begin{eqnarray}
\tilde F&=&\frac{\ep_h h_1 \tilde h_m}{\vartheta d_s (h_1+\ep_h \tilde h_m)} (u_1-\ep_u \tilde u_m)|u_1-\ep_u \tilde u_m| \nonumber \\
&=& \frac{\ep_h \tilde h_m}{\df d_s}(u_1-\ep_u \tilde u_m)|u_1-\ep_u \tilde u_m|+\mo(\ep_h^2).
\end{eqnarray}
Following the same reasoning as above, the equation \eqref{frictionP} reads:
$$
r \frac{h_m}{\df d_s}(u_1- u_m)|u_1-u_m|=gh_m\mathcal{P}.
$$
From where we obtain that 
$$
r \frac{1}{\df d_s}(u_1- u_m)^2=g\, \mathcal{P}\,\sgn(\mathcal{P})\quad \textrm{and then}\quad  u_m=v_b^{(QF)}\sqrt{\Big(\frac 1 r -1\Big)g d_s}.
$$

\section{Non-hydrostatic pressure model}\label{model_nh}

The hydrostatic hypothesis may be inaccurate and fails where non-hydrostatic pressure can affect the mobility of sediment and hence the bedload transport. We present in this section the two-layer shallow water model described in~(\ref{bilayer_model}) with a correction in the total pressure applying a similar approach to the one proposed by Yamazaki in \cite{yamazaki}, where the non-hydrostatic effects are taken into account for the nonlinear shallow water equations, hereinafter SWE.

The challenge is thus to improve nonlinear dispersive properties of the model by including information on the vertical structure of the flow while designing fast and efficient algorithms for its simulation. First we resume the development introduced by Yamazaki in \cite{yamazaki} to later apply it to our system. The idea is that in the depth averaging process, the vertical velocity average is not neglected and the total pressure is decomposed into a sum of hydrostatic and non-hydrostatic components. 

In addition, during the process of depth averaging, vertical velocity is assumed to have a linear vertical profile as well as as the non-hydrostatic pressure. Moreover, in the vertical momentum equation, the vertical advective and dissipative terms, which are small compared with their horizontal counterparts, are neglected. At the free surface we assume that the pressure vanishes as boundary condition.

Then, for a single fluid layer of thickness $h$ over a bottom $b$, the resulting $x$ and $z$ momentum equations as well as the continuity equation described in \cite{yamazaki} are

\begin{equation}
\left\{
\begin{array}{l}
\pt h +\div (h u)=0, \\ \\
\pt(h u) +\div(h u\otimes u+ h\tilde {p}_{|b+h/2})+g h \nabla_x h= - \tilde{p}_{|b}\nabla_x b,\\ \\
\pt (hw) - \tilde{p}_{|b}=0,\\ \\
\div\, u + \displaystyle\frac{w-w_{|b}}{h/2} = 0,\\
\end{array}
\right.
\label{eqn:yamazaki_original}
\end{equation}

\noindent where  $u$ is the depth averaged velocity in the $x$ direction and $w$ is the depth averaged velocity component in the $z$ direction. $\tilde {p}_{|b}$ denotes the non-hydrostatic pressure at the bottom. 
The vertical velocity at the bottom is evaluated from the seabed boundary condition
$$w_{|b} = u \nabla_x b.$$
The value of the pressure at the middle of the layer comes from the linear profile assumed for $\tilde p$ and the boundary condition $\tilde p_{|b+h}=0$, then
$$\tilde p_{|b+h/2} = \frac12\tilde p_{|b}.  $$
Note that system~(\ref{eqn:yamazaki_original}) reduces to the SWE when total pressure is assumed to be hydrostatic, and therefore $\tilde {p}_{|b}=0.$
\\\\
Now, using a similar procedure as in \cite{yamazaki}, the proposed model with non-hydrostatic effects reads: 
\begin{itemize}
	\item Water layer:
\begin{equation}
\left\lbrace
\begin{array}{l}
\pt h_1 +\div (h_1 u_1)  =0, \\ \\
\begin{split}
\pt(h_1 u_1) +\div(h_1 u_1\otimes u_1+&h_1 \tilde {p_1}_{|b+h_2+h_1/2}) \\ \\
+& g h_1 \nabla_x(b+h_1+h_2) = - \tilde{p_1}_{|b+h_2}\nabla_x(b+h_2)-F,
\end{split}\\ \\
\pt (h_1 w_1) -\tilde{p_1}_{|b+h_2}  = 0, \\ \\
\div\,  u_1+ \displaystyle\frac{w_1-{w_1}^+_{|b+h_2}}{h_1/2}  = 0. \\ \\
\end{array}
\right.
\label{eqn:water_layer}
\end{equation}
\end{itemize}
The sediment layer will be also affected by the non-hydrostatic terms. This influence comes from the boundary condition at the interface. Therefore we obtain:
\begin{itemize}
	\item Sediment layer:
\begin{equation}
\left\lbrace
\begin{array}{l}
\pt h_2 +\div (h_m u_m)=0, \\ \\
\begin{split}
\pt(h_m u_m) +&\div(h_m u_m\otimes u_m)+g h_m \nabla_x(b+rh_1+h_2)\\ \\
+&rh_m\nabla_x(\tilde{p_1}_{|b+h_2})=rF+\frac 1 2 u_m T-(1-r) gh_m \sg(u_m)\tan\delta,
\end{split}\\ \\
\pt h_f=-T, \\ \\
\end{array}
\right.
\label{eqn:sediment_layer}
\end{equation}
\end{itemize}
The system is completed with closures relations on the pressure and vertical velocity of the water layer:
$$
\tilde {p_1}_{|b+h_2+h_1/2}=\frac 1 2 \tilde {p_1}_{|b+h_2},\qquad {w_1}^+_{|b+h_2}=\pt h_2 +u_1\cdot\nabla_x(b+h_2).
$$
Note that system~(\ref{eqn:water_layer})-(\ref{eqn:sediment_layer}) reduces to~(\ref{bilayer_model}) when total pressure is hydrostatic.\\
In Section~\ref{sec:num} we will present a numerical test to show the importance of taking into account the non-hydrostatic effects in  suitable cases.

\section{Numerical scheme}\label{sec:num_scheme}
The model is solved numerically using a two-step algorithm following ideas described in~\cite{escalante}: first the hyperbolic two-layer shallow water system is solved and then, in a second step, non-hydrostatic terms will be taken into account. The source terms corresponding to friction terms are discretized semi-implicitly afterwards.

System~(\ref{eqn:water_layer}--\ref{eqn:sediment_layer}) can be written in the compact form

\begin{equation}
\left\lbrace
\begin{array}{l}
\begin{split}
\pt \sst + \px \boldsymbol{F}(\st) +& \boldsymbol{B}(\st) \px \sst = \boldsymbol{G}(\st)\px\bt \\ \\
+& \tnh(\sst, \px \sst, \bt, \px \bt, \pnh, \px \pnh) + \boldsymbol{\tau},
\end{split}\\ \\
\pt h_f = -T,\\ \\
\pt(h_1w_1) = \tilde{p_1}_{|b+h_2},\\ \\
\bnh (\sst, \px \sst, \bt, \px \bt,  w_1)=0,\\ \\
\end{array}
\right.
\label{eqn:bilayer_model_compact}
\end{equation}
where

$$
\st=\begin{bmatrix}h_1 \\q_1 \\ h_m\\q_m\end{bmatrix},\quad 
F(\st)=\begin{bmatrix}q_1 \\ \frac{q_1^2}{h_1}+\frac{1}{2}gh_1^2\\q_m\\\frac{q_m^2}{h_m}+\frac{1}{2}gh_m^2\end{bmatrix},\quad 
G(\st)=\begin{bmatrix}0 \\ -gh_1\\0\\-gh_m\end{bmatrix},\quad
\bt = b + h_f,
$$

$$ 
B(\st)=\begin{bmatrix}0 & 0& 0& 0\\ 0 & 0& gh_1& 0\\0 & 0& 0& 0\\rgh_m & 0& 0& 0\end{bmatrix}, \boldsymbol{\tau}=\begin{bmatrix}0\\ -F \\T\\rF+\frac 1 2 u_m T-(1-r) gh_m \sg(u_m)\tan\delta\end{bmatrix}.
$$
Finally, 
\begin{equation}
\tnh(\sst, \px \sst, \bt, \px \bt, \pnh, \px \pnh) = \begin{bmatrix}
0 \\ -\displaystyle\frac{1}{2}\left( h_1 \px \pnh + \pnh \px\left( h_1 +2(\bt+h_m)\right)\right) \\ 0 \\ -r h_m \px \pnh
\end{bmatrix},
\label{eqn:nht}
\end{equation}
and 
\begin{equation}
\bnh (\sst, \px \sst, \bt, \px \bt, w_1)=h_1 \px q_1 -q_1 \px\left( h_1 +2(\bt+h_m)\right) + 2h_1w_1 + 2h_1\px{q_m}.
\label{eqn:elipt}
\end{equation}
Note that last equation in (\ref{eqn:water_layer}), corresponding to free-divergence equation, has been multiplied by $h_1$, giving equation~(\ref{eqn:elipt}).

We describe now the numerical scheme used to discretize the 1D system~(\ref{eqn:bilayer_model_compact}). We shall omit the discretization of the friction and mass-transference terms and it will be applied in a semi-implicit manner. In a first step, we shall solve the hyperbolic problem. Then, in a second step, non-hydrostatic terms will be taken into account.

\subsection{Finite volume discretization for the two-layer SWE}

The two-layer SWE written in vector conservative form is given by

\begin{equation}
\pt \sst + \px \boldsymbol{F}(\st) + \boldsymbol{B}(\st) \px \sst = \boldsymbol{G}(\st)\px\bt.
\label{eqn:2layersw}
\end{equation}

The system is solved numerically by using a finite volume method. As usual, we subdivide the horizontal spatial domain into standard computational cells $I_i=[x_{i-1/2},x_{i+1/2}]$ with length $\Delta x_i$ and define 

$$ \sst_i(t) = \frac{1}{\Delta x_i}\int_{I_i} \sst(x, t) dx,$$
the cell average of the function $\sst(x, t)$ on cell $I_i$ at time $t$.
We shall also denote by $x_i$ the center of the cell $I_i$. For the sake of simplicity, let us assume that all cells have the same length $\Delta x.$

\begin{figure}
	\centering
	\includegraphics[scale=1.]{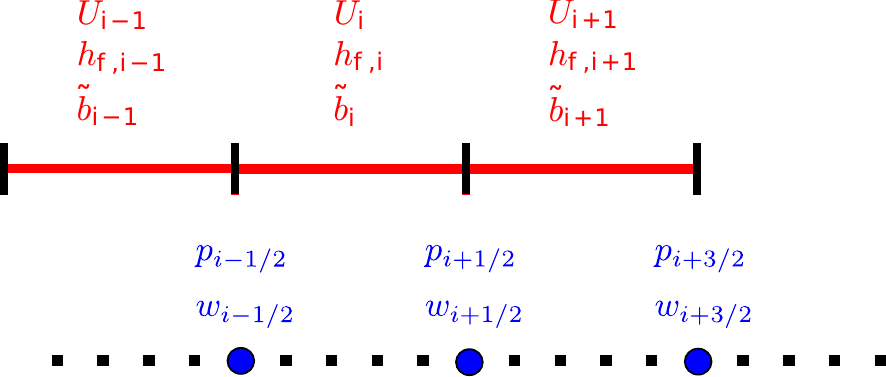}
	\caption{Numerical scheme stencil. Up: finite volume mesh. Down: staggered mesh for finite differences.}
	\label{fig:stencil}
\end{figure}

In~\cite{edofer2011} authors introduce a first order path-conservative numerical scheme, named IFCP, which is constructed by using a suitable decomposition of a Roe matrix of system~(\ref{eqn:2layersw}) by means of a parabolic viscosity matrix, that captures information of the intermediate fields. IFCP is a path-conservative scheme in the sense defined in~\cite{pares2006}. IFCP numerical scheme can be written as follows

$$ \st_i^{n+1} = \st_i^n - \frac{\Delta t}{\Delta x}\left( D_{i-1/2}^+ + D_{i+1/2}^- \right), $$
being $D_{i-1/2}^\pm = D_{i-1/2}^\pm(\st_i, \st_{i+1}, \bt_i, \bt_{i+1})$ defined by 

\begin{equation} \label{def_d_p_m}
\begin{array}{ll}
D^{\pm}_{i+1/2} = &   {\displaystyle \frac{1}{2}\left( F(\st_{i+1}^n)- F(\st_i^n)+B_{i+1/2}(\st_{i+1}^n-\st_i^n) \right.} \\
& -G_{i+1/2}(\tilde{b}_{i+1}-\tilde{b}_i)  \\
& \pm \left. Q_{i+1/2}\left(\st_{i+1}^n-\st_i^n -A_{i+1/2}^{-1}G_{i+1/2}(\tilde{b}_{i+1}-\tilde{b}_i)\right)\right),
\end{array}
\end{equation}
where
$$
B_{i+1/2} = \begin{bmatrix}0 & 0& 0& 0\\ 0 & 0& gh_{1,i+1/2}& 0\\0 & 0& 0& 0\\rgh_{m,i+1/2} & 0& 0& 0\end{bmatrix},\quad
G_{i+1/2}=\begin{bmatrix}0 \\ -gh_{1,i+1/2}\\0\\-gh_{m,i+1/2}\end{bmatrix},
$$
$$
A_{i+1/2} = J_{i+1/2} + B_{i+1/2},
$$
being $J_{i+1/2}$ a Roe linearization of the Jacobian of the flux $F$ in the usual sense:
$$
A_{i+1/2} = \begin{bmatrix}0 & 1& 0& 0\\ -u_{1,i+1/2}^2+gh_{1,i+1/2} & 2u_{1,i+1/2}& gh_{1,i+1/2}& 0\\0 & 0& 0& 1\\rgh_{m,i+1/2} & 0& -u_{m,i+1/2}^2+gh_{m,i+1/2}& 2u_{m,i+1/2}\end{bmatrix},
$$
and
$$ h_{*, i+1/2}=\displaystyle\frac{h_{*,i}+h_{*,i+1}}{2},\ u_{*, i+1/2} = \displaystyle\frac{u_{*, i}\sqrt{h_{*,i}}+u_{*,i+1}\sqrt{h_{*,i+1}}}{\sqrt{h_{*,i}} + \sqrt{h_{*,i+1}}},\ *\in{\{ 1,2\}}.$$
The key point is the definition of the matrix $Q_{i+1/2 }$ ,that in the case of the IFCP is defined by:
$$Q_{i+1/2} = \alpha_0 Id + \alpha_1 A_{i+1/2} + \alpha_2 A_{i+1/2}^2,$$
where $\alpha_j,\ j=0,1,2$ are defined by:
$$
\begin{bmatrix}1 & \lambda_1& \lambda_1^2\\ 1 & \lambda_4& \lambda_4^2\\1 & \chi_{int}& \chi_{int}^2&\end{bmatrix}
\begin{bmatrix}\alpha_0 \\ \alpha_1 \\ \alpha_2\end{bmatrix}=
\begin{bmatrix}|\lambda_1| \\ |\lambda_4| \\ |\chi_{int}|\end{bmatrix},
$$
where
$$ \chi_{int} = \mathcal{S}_{ext}\max(|\lambda_2|,|\lambda_3|), $$
with
$$
\mathcal{S}_{ext}=\left\{\begin{array}{ll}
\textrm{sgn}(\lambda_1+\lambda_4), &  \mbox{if } (\lambda_1+\lambda_4)\neq 0, \\ \\
1, &  \textrm{otherwise}. \\
\end{array}\right.
$$
It can be proved that the numerical scheme is linearly~$L^\infty$~--stable under the usual CFL condition.
\subsection{Treatment of the non-hydrostatic terms}
Regarding non-hydrostatic terms, we consider a staggered-grid (see Figure~\ref{fig:stencil}) formed by the points $x_{i-1/2},\ x_{i+1/2}$ of the interfaces for each cell $I_i$, and denote the point values of the functions $\pnh$ and $w_1$ on point $x_{i+1/2}$ at time $t$ by
\[p_{i+1/2}(t)=\pnh(x_{i+1/2},t),\ w_{i+1/2}(t)=w_1(x_{i+1/2},t).\]

Following~\cite{escalante}, $\pnh$ and $w_1$ will be discretized using second order compact finite differences. In order to obtain point value approximations for the non-hydrostatic variables $p_{i+1/2}$ and $\ w_{i+1/2},$ operator $\bnh (\sst, \px \sst, \bt, \px \bt, w_1)$ will be approximated for every point $x_{i+1/2}$ of the staggered-grid (Figure~(\ref{fig:stencil})). Then, a second order compact finite-difference scheme is applied to 

\begin{equation}
\left\lbrace
\begin{array}{l}
\pt \sst = \tnh(\sst, \px \sst, \bt, \px \bt, \pnh, \px \pnh),\\ \\
\pt(hw_1) = \tilde{p_1}_{|b+h_2},\\ \\
\bnh (\sst, \px \sst, \bt, \px \bt, w_1)=0,\\ \\
\end{array}
\right.
\label{eqn:bilayer_model}
\end{equation}
where the values obtained in previous step are used as initial condition for the system. The resulting linear system is solved using an efficient Thomas algorithm.

\section{Numerical tests}\label{sec:num}

In this section we present three numerical tests, for the  model proposed in section~\ref{model_nh} with the quadratic friction law proposed in~(\ref{fric_q}). The first one corresponds to the evolution of a dune, where the computed velocity of the two-layer model and the one deduced for the SVE model are compared.  The second test is a dam break problem over an erodible sediment layer where laboratory data is used to validate the model. The test proves as well its validity for regions where the interaction between the fluid and the sediment is strong. In those situations the velocities computed by the two-layer model and the SVE one are not close. The last test shows the difference between the hydrostatic and the non-hydrostatic model on shape of the bed surface.

The numerical results follow from a combination of the scheme described in Section~\ref{sec:num_scheme} with a discrete approximation of bottom and surface derivatives. The numerical simulations are done with a CFL number equal to 0.9.

\subsection{Test 1: bedload transport}
\begin{figure}
	\begin{center}
		\includegraphics[width=0.9\textwidth]{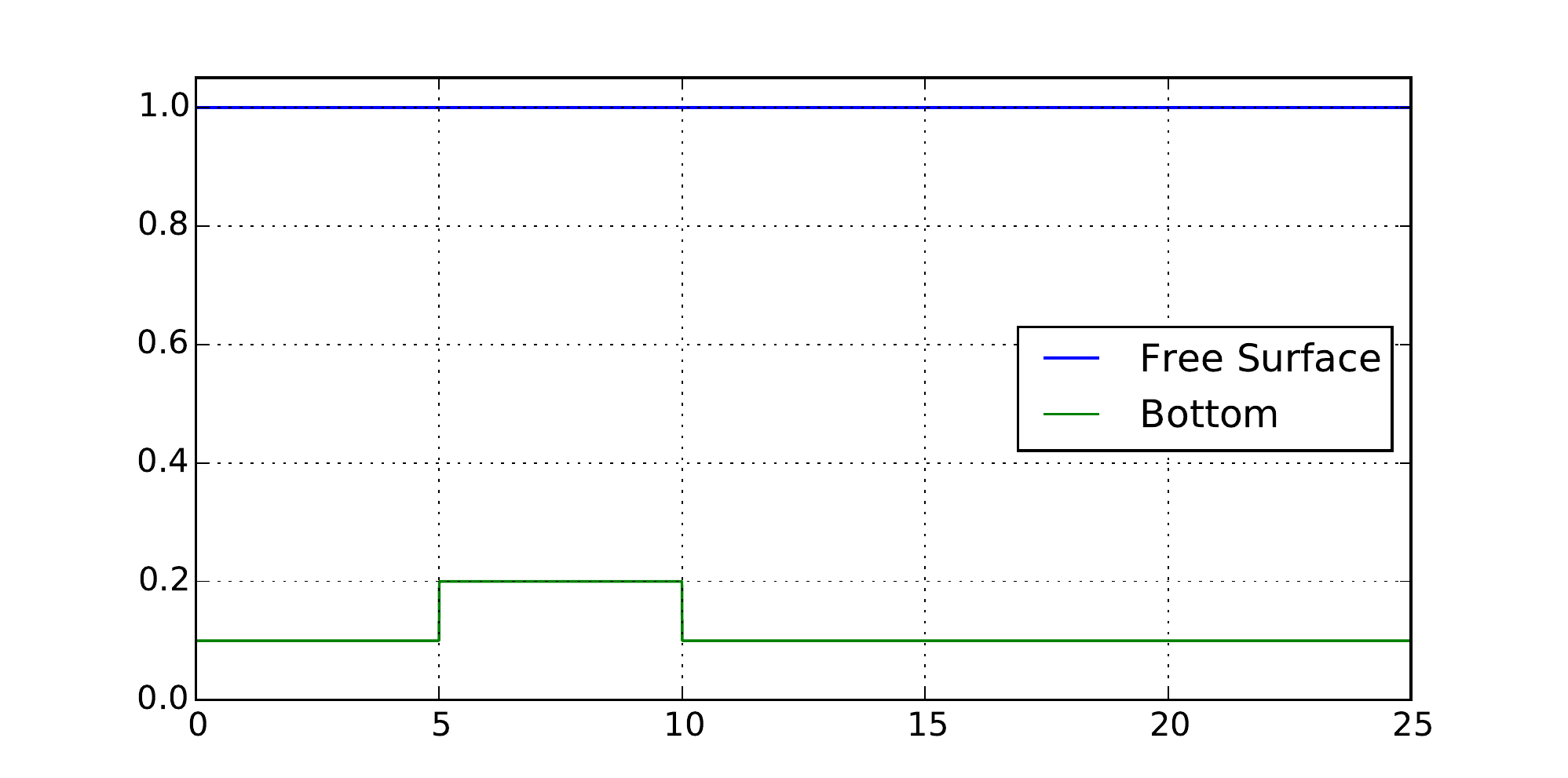}
	\end{center}
	\caption{Test 1: Initial condition}
	\label{fig:dunaSW_t0}
\end{figure}

\begin{figure}
	\begin{center}
		\includegraphics[width=0.49\textwidth]{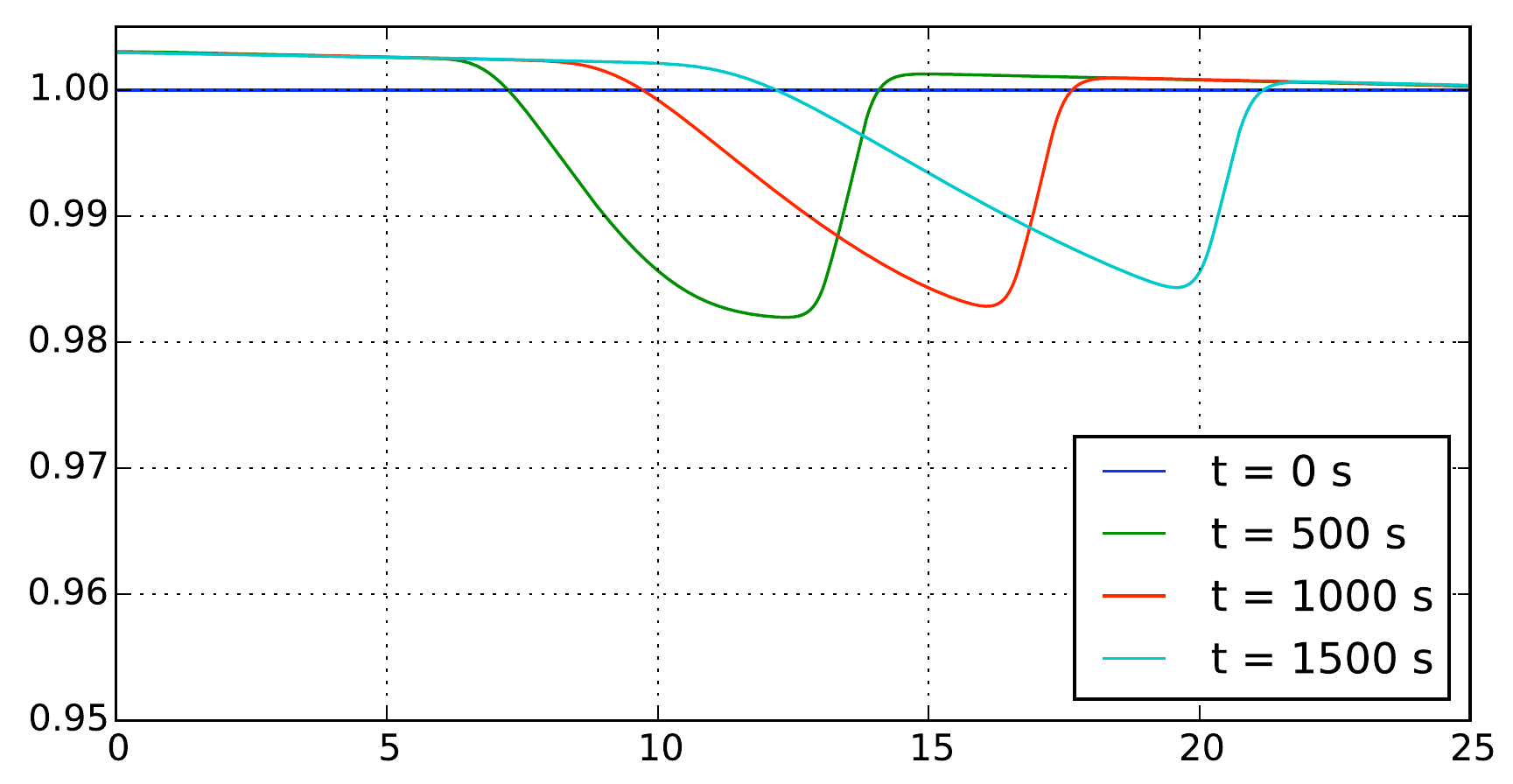}
		\includegraphics[width=0.49\textwidth]{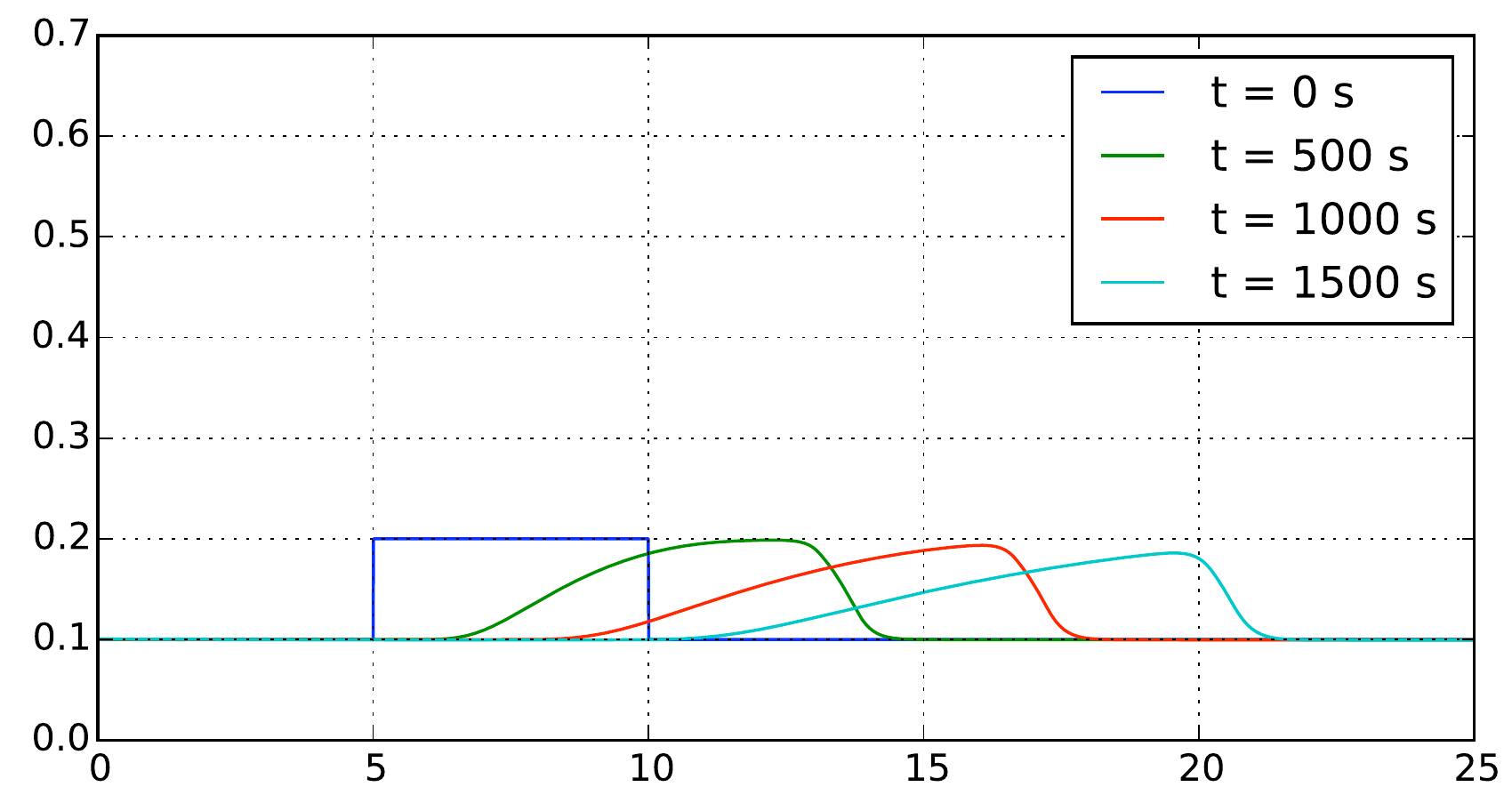}
	\end{center}
	\caption{Test 1: (a) Surface  and (b) bottom, at times $t=0,\ 500,\ 1000,\ 1500\ s$}
	\label{fig:dunaSW_superficies}
\end{figure}

\begin{figure}[!h]
	\begin{center}
		\includegraphics[width=0.49\textwidth]{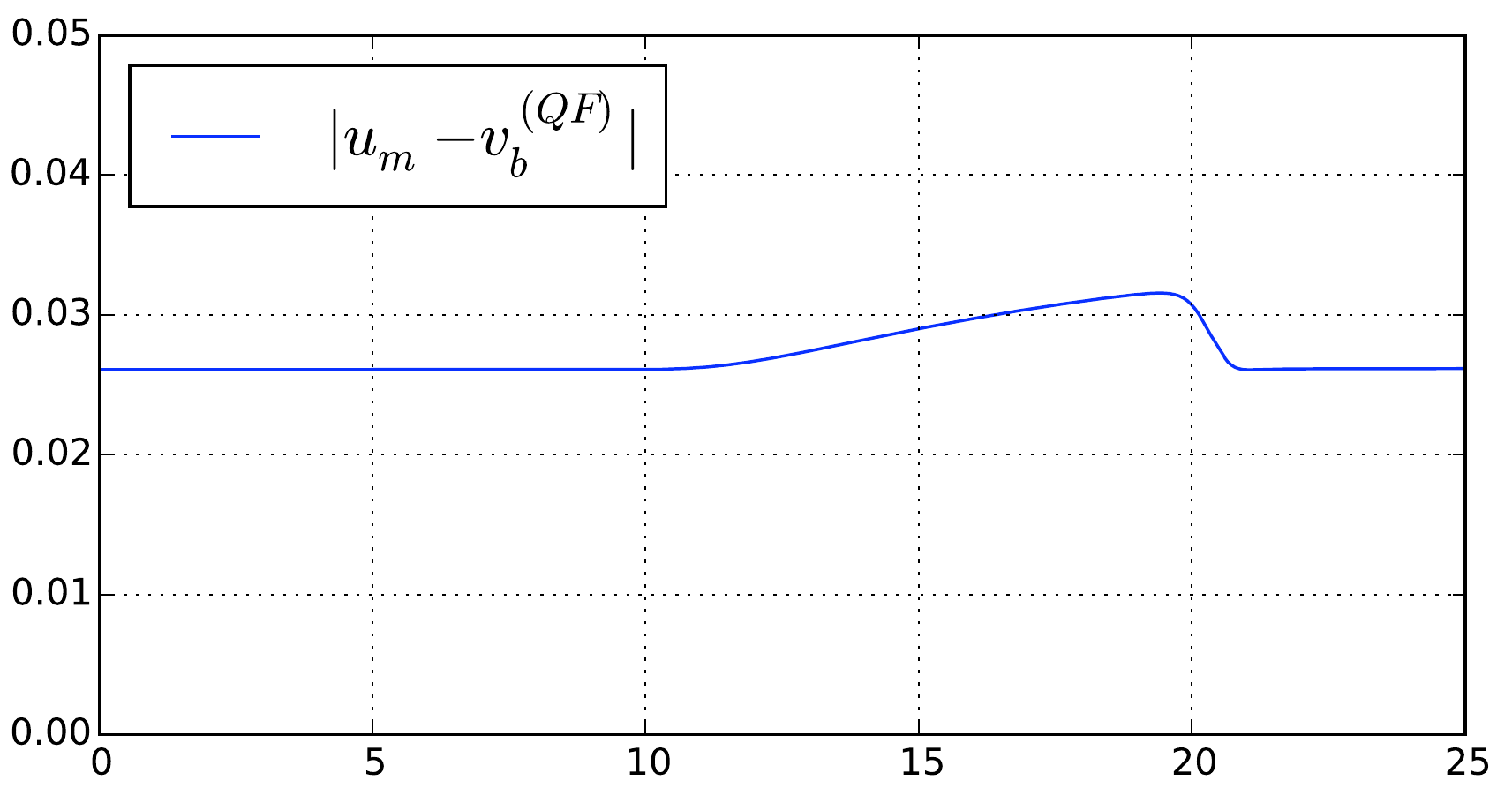}
		\includegraphics[width=0.49\textwidth]{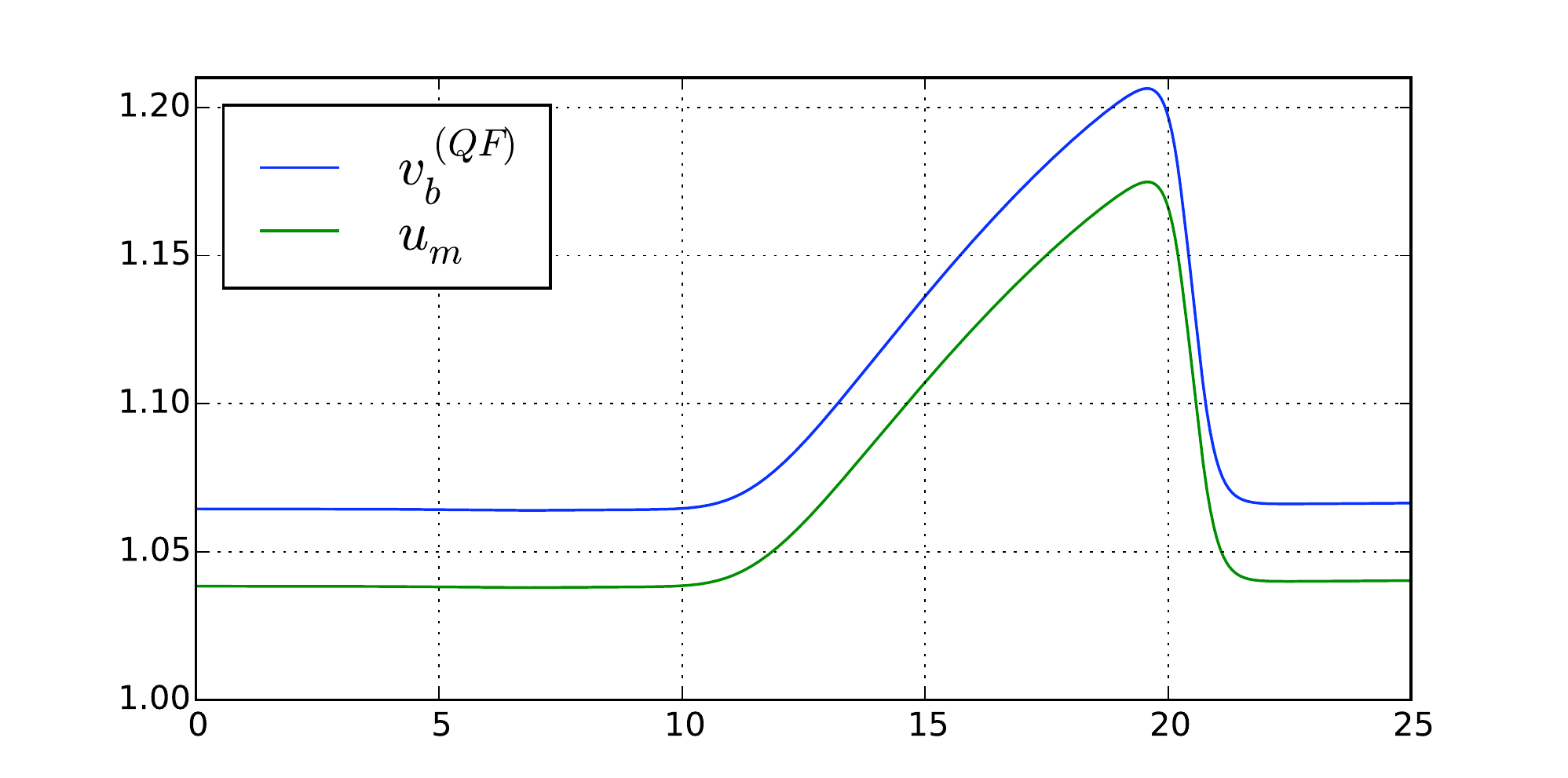}
	\end{center}
	\caption{Test 1: Comparison between $u_m$ and $v_b^{(QF)}$ at time $t=1500\ s$ }
	\label{fig:duna_difu2u2star}
\end{figure}

\begin{figure}[!h]
	\begin{center}
		\includegraphics[width=0.7\textwidth]{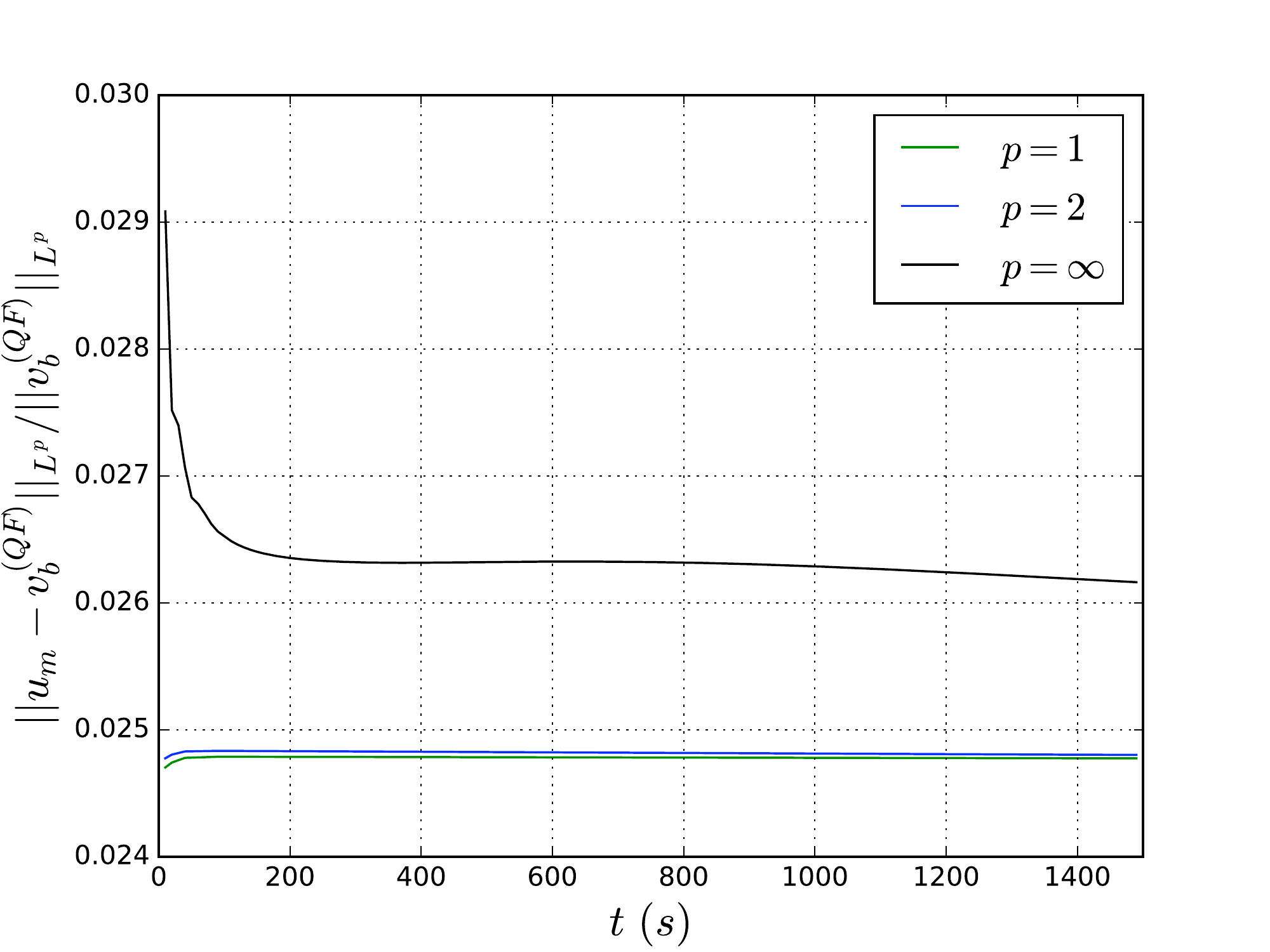}
	\end{center}
	\caption{Test 1: Evolution in time of the relative error between $u_m$ and $v_b^{(QF)}$ in  $L^1$, $L^2$ and $L^{\infty}$ norm. }
	\label{fig:duna_absoluto_relative_error}
\end{figure}

In this test we would like to show the ability of the proposed model to reproduce the bedload transport. In particular we study the formation and evolution of a dune. To do so, let us consider the following initial condition over a domain of 25m, (see Figure~{\ref{fig:dunaSW_t0}}):
\begin{equation*}
h_2(0,x)=\left\{
\begin{array}{ll}
0.2\text{ m}, &\mbox{ if }x\in[5,10], \\
0.1\text{ m}, &\mbox{otherwise.}
\end{array}
\right. \quad h_1(0,x)+h_2(0,x)=1\text{ m}, \quad q_1(0,x)=1 \text{ m}^2/\text{s}^2.
\end{equation*}
The fixed bottom is set to $b(x)=0$. We set left boundary condition $q_1(t,0)=1\text{m}^2/\text{s}^2,$ and open boundary condition on the right hand side.
The parameters for the model in this academic test have been set as follows: 
$ r=0.34, \quad d_s=0.01\text{m}, \quad \theta_c = 0.047, \quad \delta=25^o.$
Additionally for the transference term we introduce:
$ K_e = 0.1,\ K_d = 0.01, \quad \varphi=0.4,\quad n = 0.01.$
We use a discretization of 5000 points for the computational domain.\\\\ 
In Figure {\ref{fig:dunaSW_superficies}} we show the free surface and the sediment bottom surface at different times.
We can see the dune profile that is transported by the flow.
\\
In Figure \ref{fig:duna_difu2u2star} we show the difference between the velocities $u_m$ obtained by the model and $v_b^{(QF)}$, the velocity deduced in \cite{FerMoNarZab} for the Saint Venant Exner model given in equation (\ref{vbqf}). The difference is of  order $10^{-2}$ at last computed time $t=1500$s. In Figure \ref{fig:duna_absoluto_relative_error} we show the evolution in time of the relative error between $v_b^{(QF)}$ and $u_m$. We remark that it remains constant in time and the difference is small. The results show that the model behaves well in low bedload transport regimes and behaves in a similar way as a SVE would.

\FloatBarrier

\subsection{Test 2: comparison with experimental data}

The purpose of this second test is to validate the model with experimental data. This experiment has been realized in the laboratory by Spinewine - Zech \cite{spinewine_test}. The same numerical test has been performed in \cite{ung_tesis} using classical models. Nevertheless we remark again that classical models make and assumption of low transport regime so that this test is not in the range of their validity. 

\begin{figure}[!h] 
\begin{center}
\includegraphics[width=0.7\textwidth]{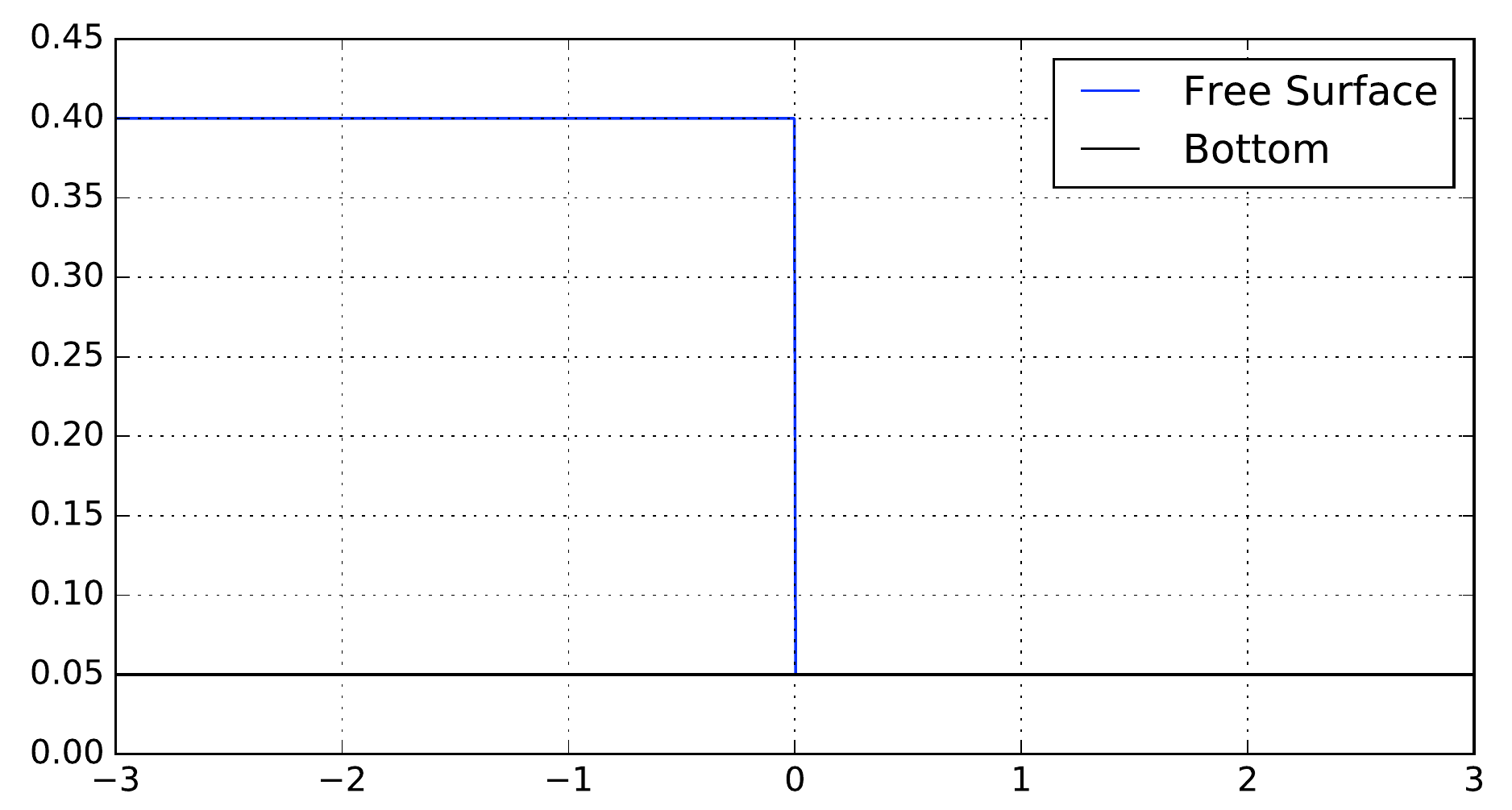}
\end{center}
\caption{Test 2: Initial condition}
\label{fig:presaSW_t0}
\end{figure}

\noindent
The experimental test takes place in a rectangular flume, the river bed is flat and at the initial state, the retained water mass is released. To do so, let us consider the following initial condition at the domain $[-3\text{ m},3\text{ m}]$ (see Figure~{\ref{fig:presaSW_t0}}):
\begin{equation*}
h_1(0,x)=\left\{
\begin{array}{ll}
10^{-12}\text{ m}, &\mbox{ if }x>0, \\
0.35\text{ m}, &\mbox{ if }x\leq 0,
\end{array}
\right. \quad h_2(0,x)=0.05\text{ m}, \quad q_1(0,x)=q_2(0,x)=0 \text{ m}^2/\text{s}^2.
\end{equation*} 
The fixed bottom is set to zero and we set open boundary conditions. We choose the parameters given in \cite{ung_tesis},
$$ r=0.63, \quad d_s=0.0039\text{ m}, \quad \theta_c = 0.047, \quad \delta=35^o,$$
$$ K_e = 0.1,\quad K_d = 0.15, \quad \varphi=0.4,, \quad n = 0.0039.$$

\begin{figure}[!h]
\begin{center}
\includegraphics[width=0.9\textwidth]{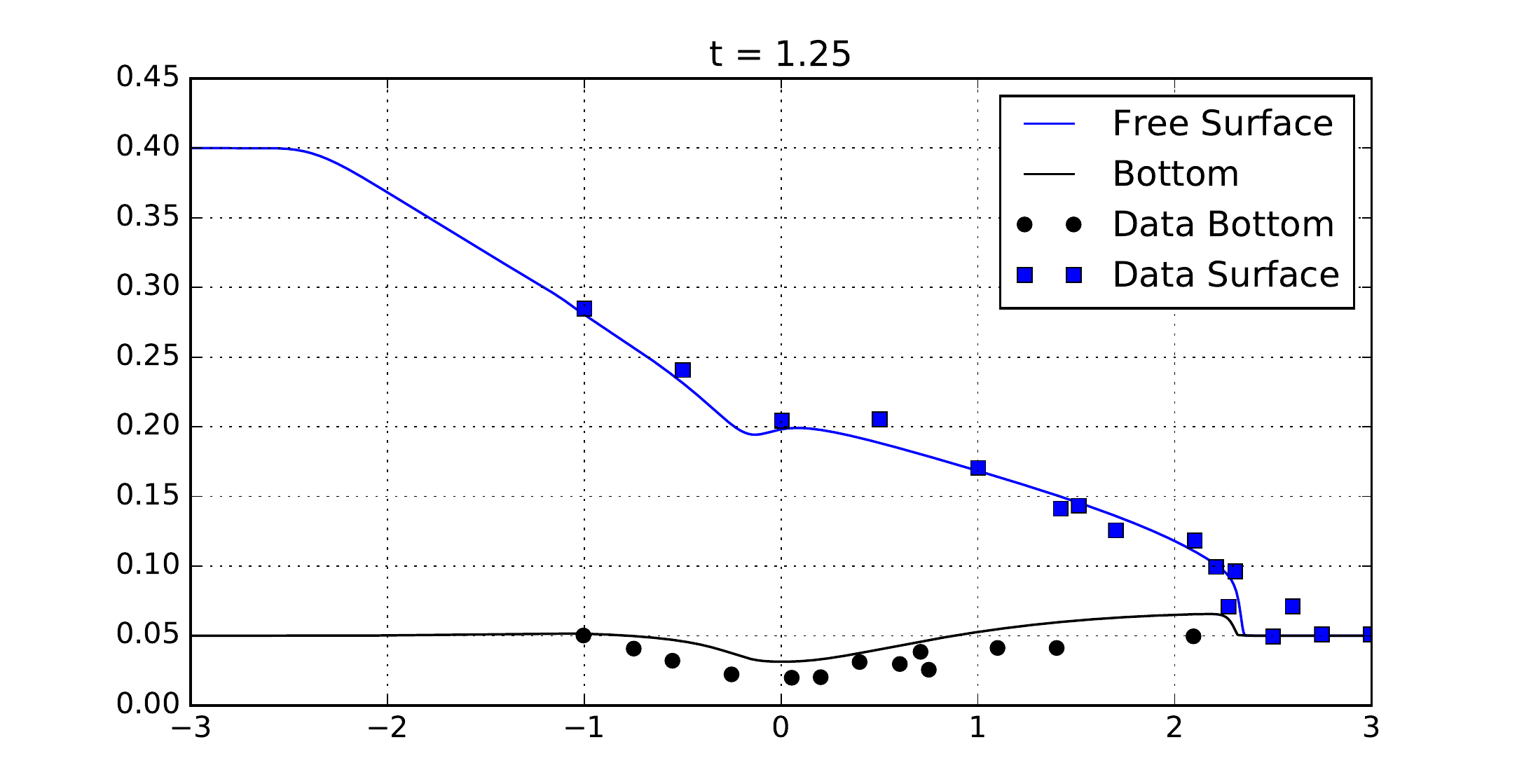}
\end{center}
\caption{Test 2: Comparison with experimental data at time $t=1.25\ s$}
\label{fig:presaSW_comparison}
\end{figure}

\begin{figure}[!h]
	\begin{center}
		\includegraphics[width=0.49\textwidth]{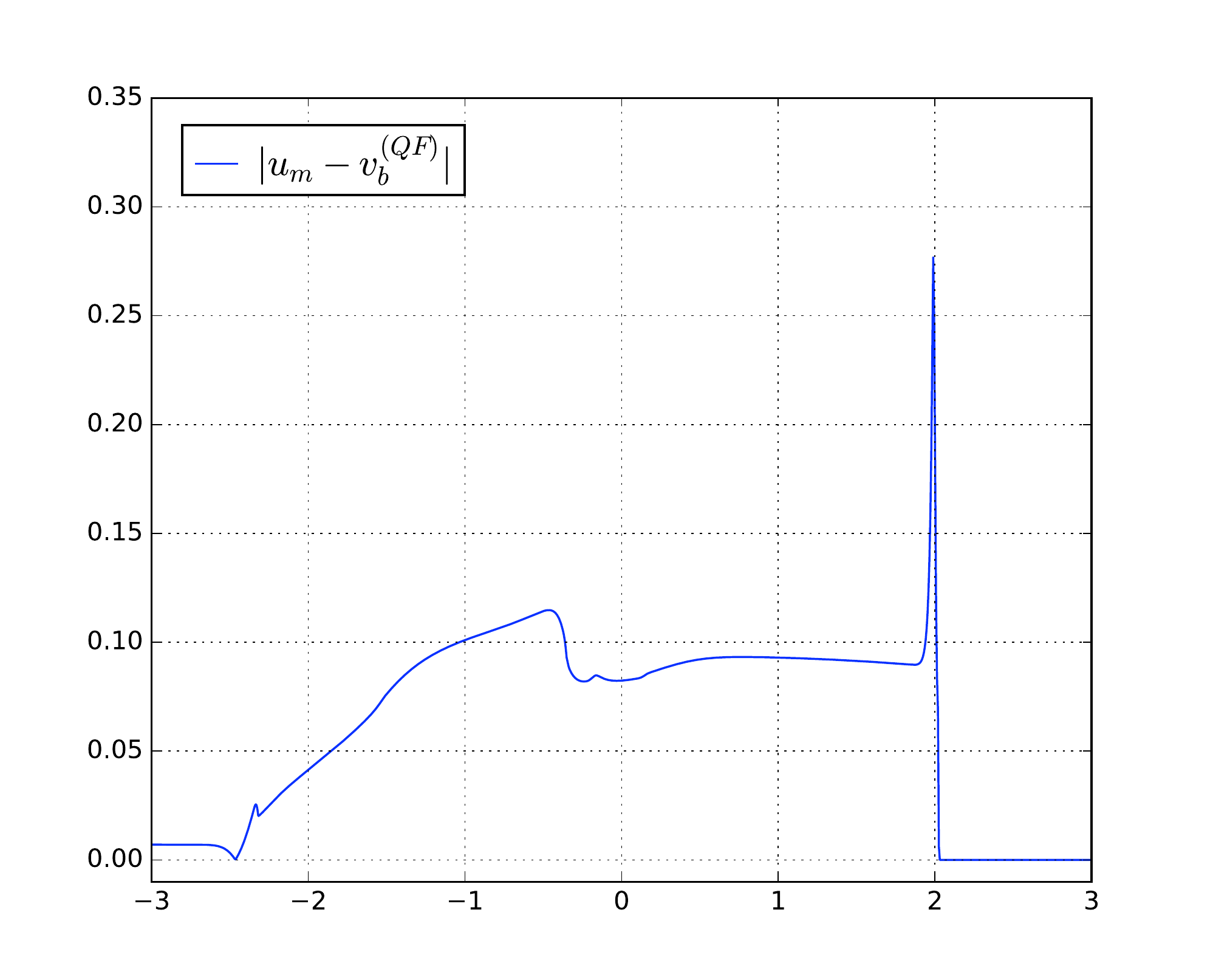}
		\includegraphics[width=0.48\textwidth]{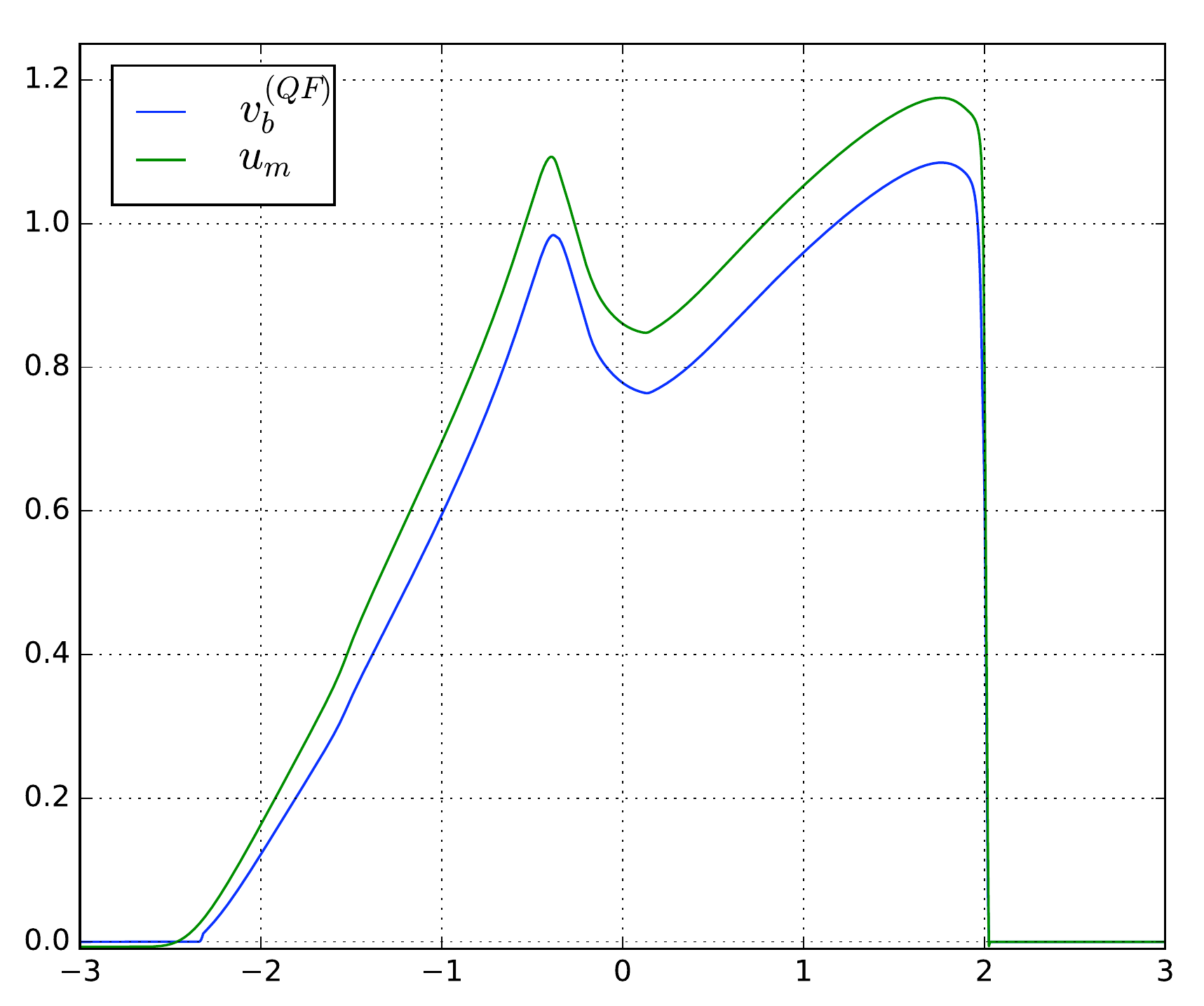}
	\end{center}
	\caption{Test 2: Difference between $u_m$ and $v_b^{(QF)}$ at time $t=1.25\ s$ }
	\label{fig:presa_difu2u2star}
\end{figure}

\begin{figure}[!h]
	\begin{center}
		\includegraphics[width=0.7\textwidth]{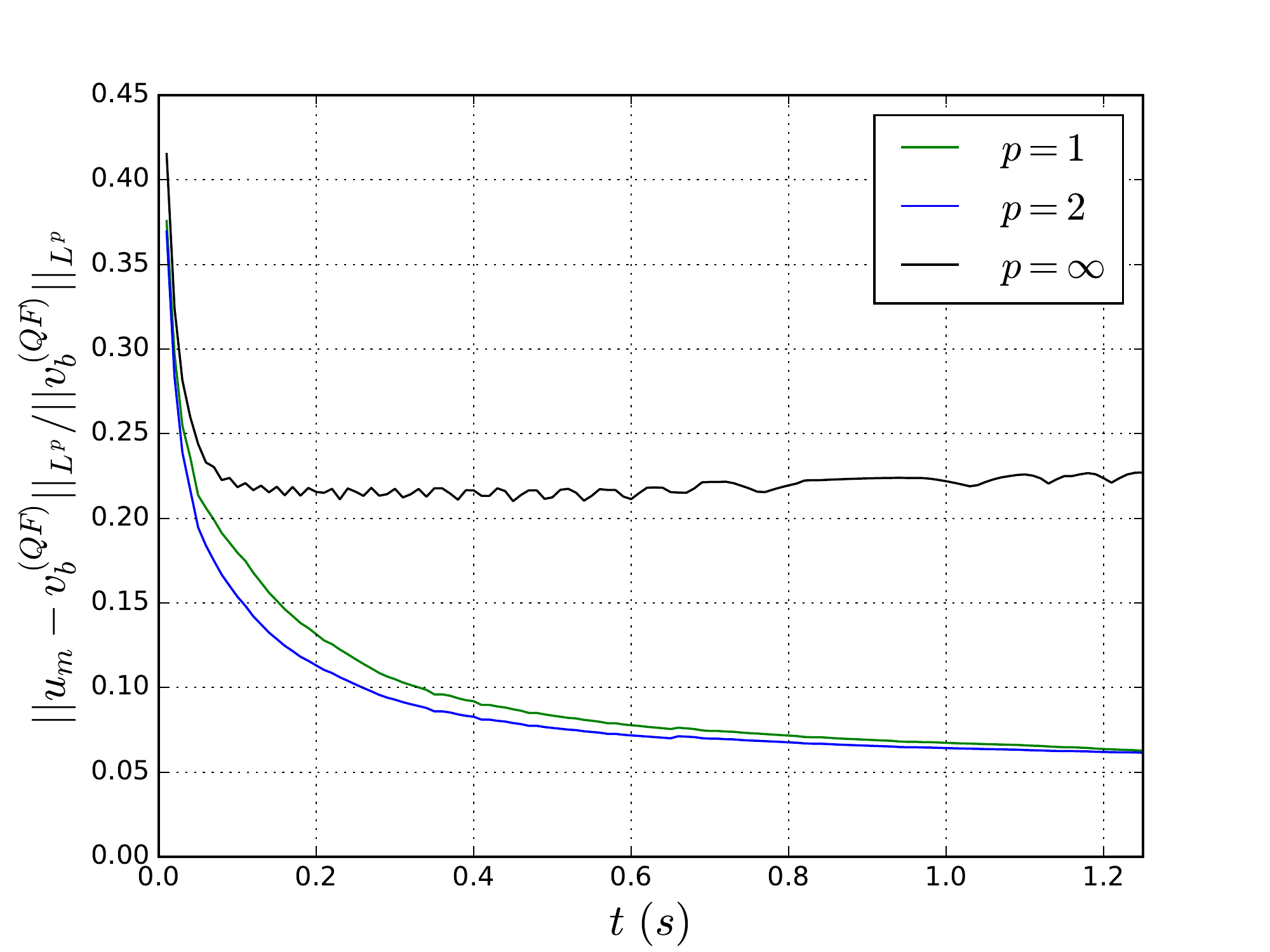}
	\end{center}
	\caption{Test 2: Evolution in time of the relative error between $u_m$ and $v_b^{(QF)}$ in $L^1$, $L^2$ and $L^{\infty}$ norm. }
	\label{fig:presa_absoluto_relative_error}
\end{figure}

\noindent
The computational domain used is discretized with $1000$ points.\\
In Figure~{\ref{fig:presaSW_comparison}} we show the free surface and the sediment bottom surface at time $t=1.25\ s$ compared with experimental data.
Both free surface and sediment bottom are captured successfully. Comparing with the results provided in~\cite{ung_tesis} with classical models, we observe a better fit to the data for the advancing front of the sediment. 
\\
As expected, in this test the difference between the velocities $u_m$ and $v_b^{(QF)}$ is not so small (see Figure \ref{fig:presa_difu2u2star}). Thus, the hypothesis of sediment transport in a small morphodynamic time is not suitable for this test. In Figure \ref{fig:presa_absoluto_relative_error} we show the evolution in time of the relative error between $v_b^{(QF)}$ and $u_m$. We can observe that the error in norm $L^{\infty}$ remains constant until the simulated time, where the error in norms $L^1$ and $L^2$ decrease. Nevertheless, during all the simulation these errors are of a degree of magnitude bigger than for the case of a dune evolution presented in previous test. The difference is specially relevant at the advancing front, where the hypothesis of low transport rate is no longer valid. The new model introduced here does not requires such assumption.

\clearpage

\subsection{Test 3: non-hydrostatic effects}

The focus now is to study the influence of the non-hydrostatic effects on the sediment layer, solving the model proposed in section~\ref{model_nh}.  The computational domain is $[0\text{ m},15\text{ m}]$. Let us consider the following initial condition
\begin{equation*}
h_1(0,x)=0.8\text{ m} \quad h_2(0,x)=0.2\text{ m}, \quad q_1(0,x)=q_2(0,x)=0\text{ m}^2/\text{s}^2.
\end{equation*}

\begin{figure}[!h]
	\centering
	\subfigure[Non-hydrostatic pressure model]{
		\includegraphics[width=0.47\textwidth]{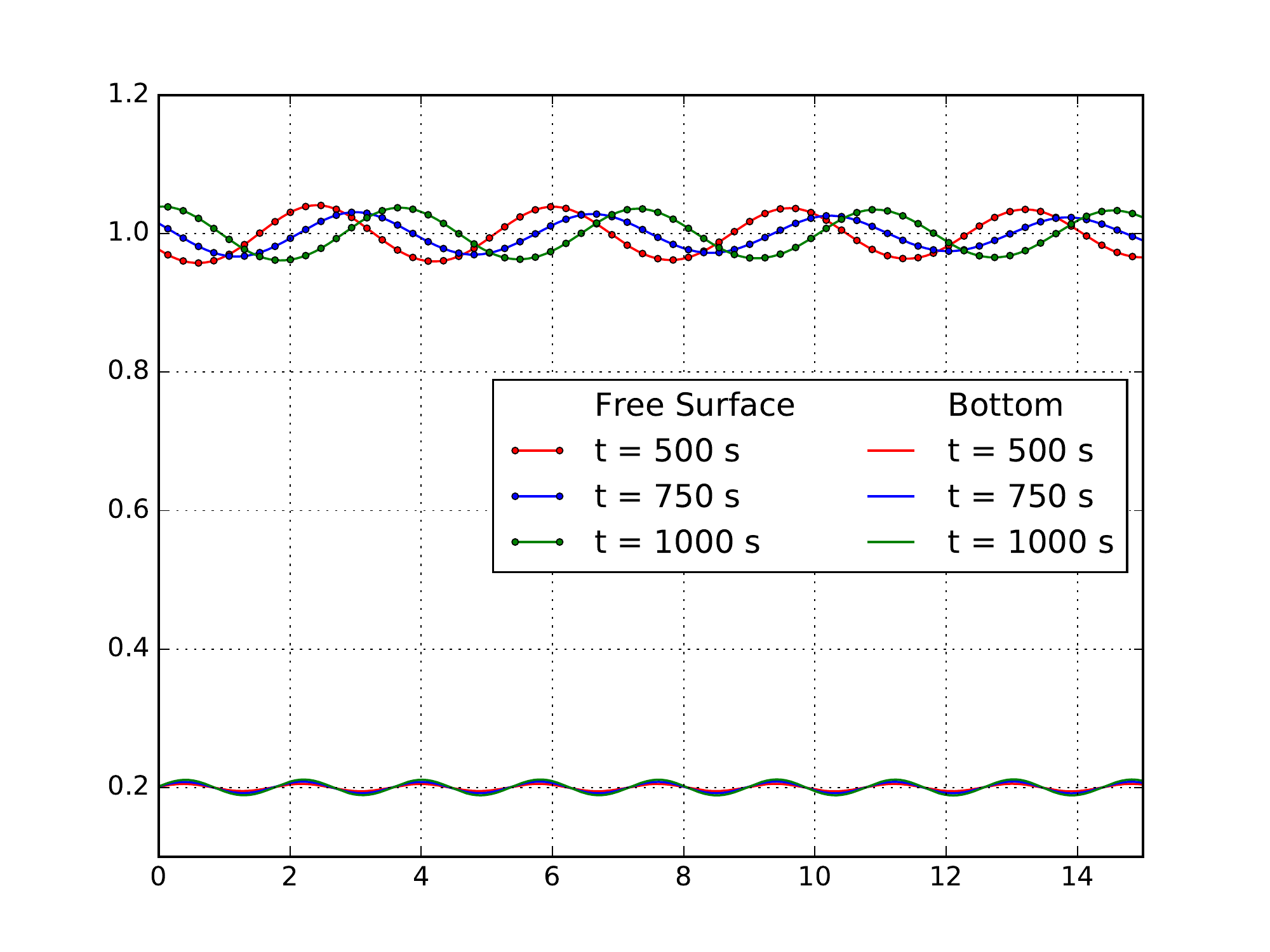}
		\label{fig:pattern_times_NH}
	}
	\subfigure[Hydrostatic pressure model]{
		\includegraphics[width=0.47\textwidth]{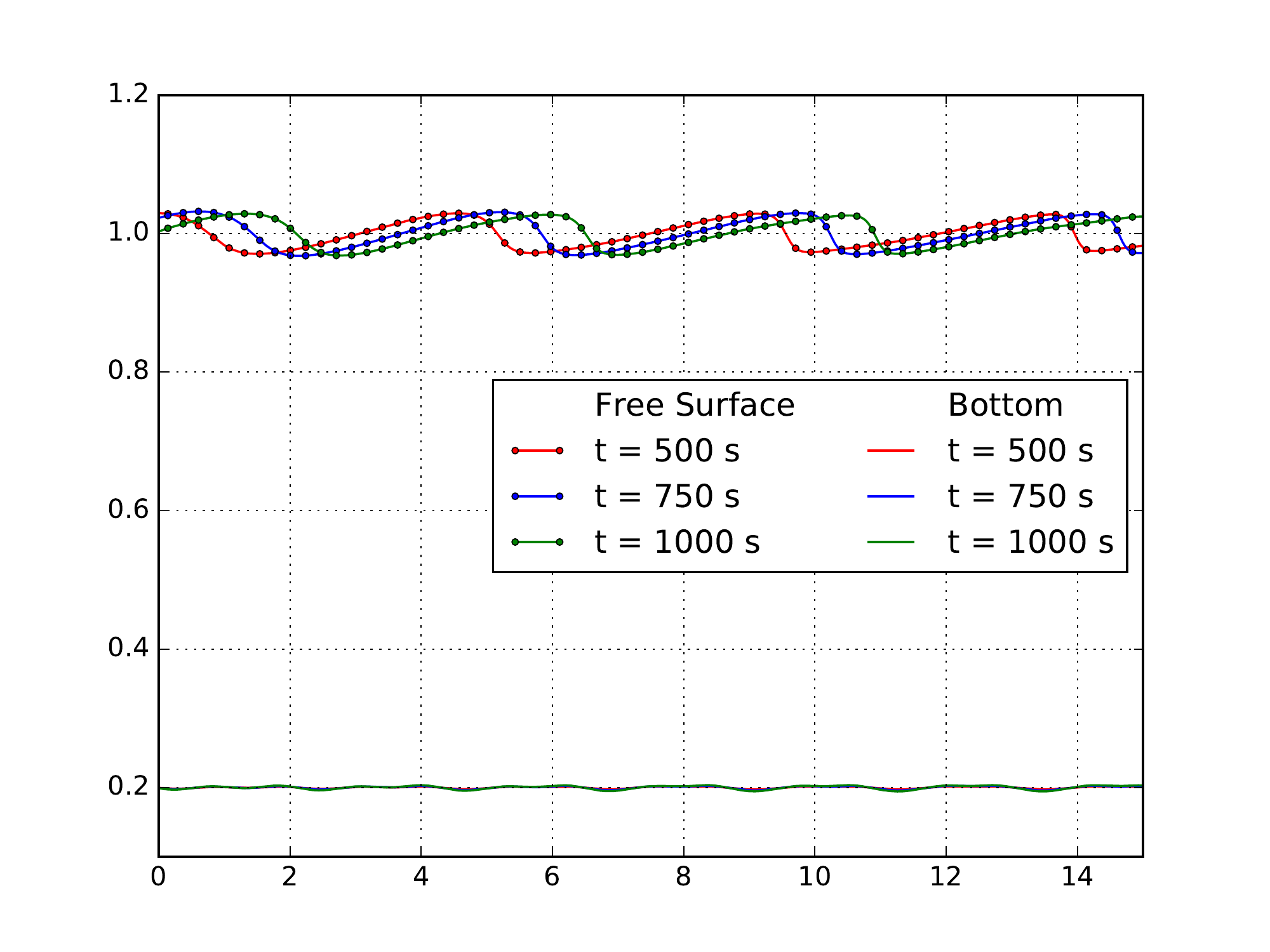}
		\label{fig:pattern_times_NH}
	}
	\caption[Optional caption for list of figures]{Test 3. Evolution of the sediment bottom surface at times $t=250,\ 500,\ 750,\ 1000\ s$ for the hydrostatic model \eqref{bilayer_model} and non-hydrostatic model \eqref{eqn:water_layer}-\eqref{eqn:sediment_layer}}
	\label{fig:pattern_times}
\end{figure}

\begin{figure}[!h]
	\centering
	\subfigure[Free surface at time $t= 1000\ s$]{
		\includegraphics[width=7.5cm]{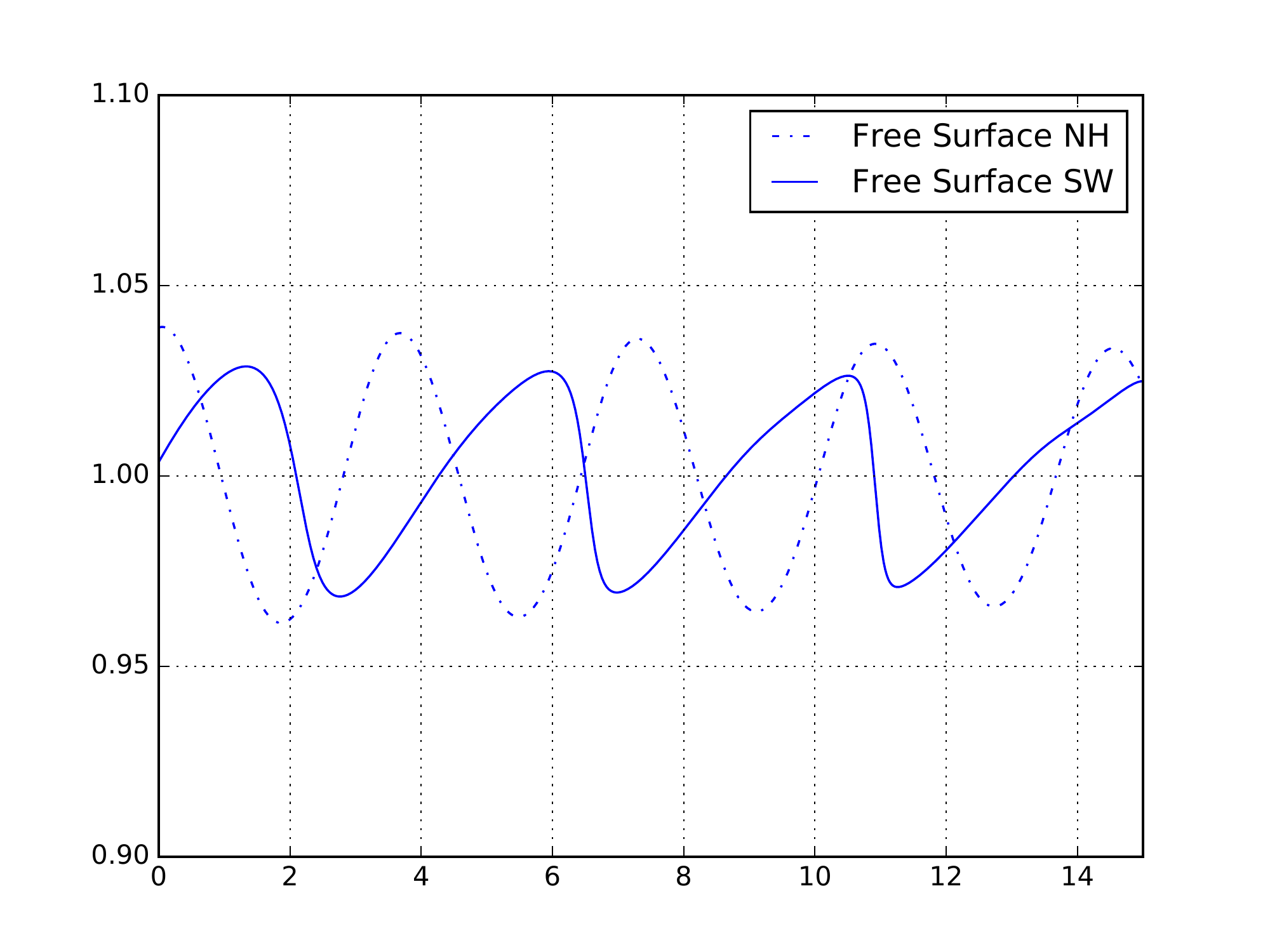}
		\label{fig:pattern_comparison_surface}
	}
	\subfigure[Bottom at time $t= 1000\ s$]{
		\includegraphics[width=7.5cm]{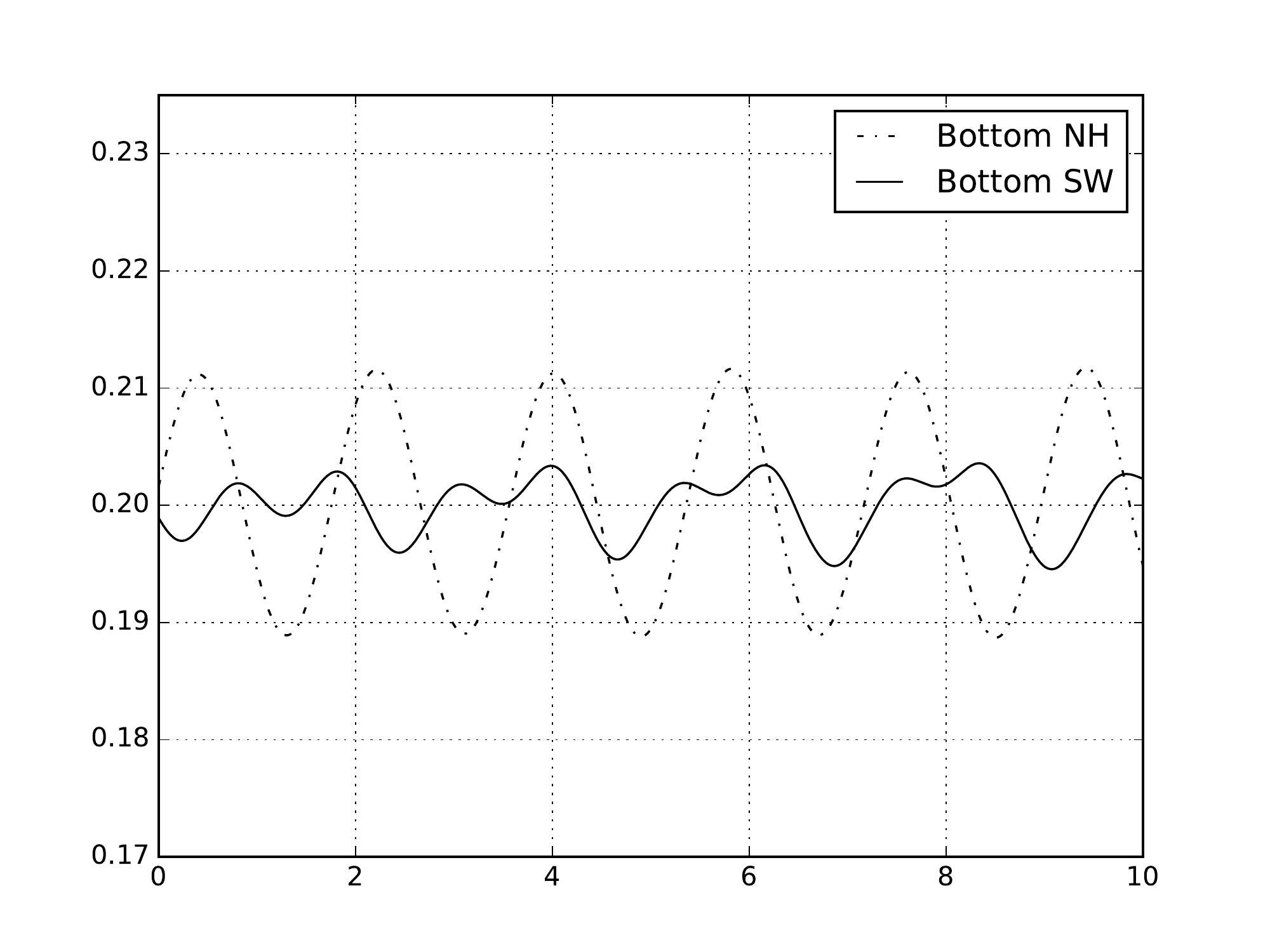}
		\label{fig:pattern_comparison_bottom}
	}
	\caption[Optional caption for list of figures]{Test 3. Comparison of the free surface \subref{fig:pattern_comparison_surface} and the bottom \subref{fig:pattern_comparison_bottom} at time $t=1000\ s$ for the hydrostatic (SW) and non-hydrostatic (NH) model}
	\label{fig:pattern_comparison}
\end{figure}

\noindent 
An incoming wave train is simulated through the left boundary condition $h_1(t,0)=0.8 + 0.1\sin(5t),$ and open boundary condition on the right hand side is considered.
The parameters for the model have been set as follows $ r=0.63, \quad d_s=0.1, \quad \theta_c = 0.047, \quad \delta=35^o,$ $ K_e = 0.001,\quad K_d = 0.01, \quad \varphi=0.4,\quad n=0.1.$
We take $1200$ points to discretize the domain.\\
In Figure~{\ref{fig:pattern_times}} we show the evolution of the sediment bottom surface at times $t=500,\ 750,\ 1000\ s$ for the model with and without non-hydrostatic effects. A more detailed comparison of the final bottom and free surface can be seen in Figure~{\ref{fig:pattern_comparison}}. Freds{\o}e and Deigaard~ \cite{fredsoe1992mechanics} described the behaviour of finite amplitude dunes under a unidirectional current. As it can be seen, the pattern wave generated by taking into account the non-hydrostatic effects, is much more similar to the pattern that can be observed in nature, close to coastal areas.
\label{key}

\FloatBarrier

\section{Conclusions}
A two-layer shallow water model has been presented that can deal with intense and slow bedload sediment transport. A version that includes non-hydrostatic pressure in the fluid layer and its influence on the gradient pressure of the sediment layer is also considered. In the case that the velocity of the fluid and the sediment layer are very different we are in a regime where SVE models can be considered. In this case we show that the proposed two-layer model converges to the SVE with gravitational effects. In the numerical tests we have shown that the velocities of the two-layer and the SVE models are very close in the case of the evolution of a dune. Nevertheless, its difference can become  relevant in the case of a dam break problem. In that situation we have also compared with experimental data with good results. Finally, in the numerical test section we have shown that the non-hydrostatic effects become important in the bed form induced by sinusoidal waves. 

From a computational point of view the main difference between  considering a SVE model or the two-layer system is the inclusion of gravitational effects without any extra computational cost. In the SVE model it is commonly included by considering the effective Shields parameter. Although we have shown in the introduction of the paper that the usual definition of the effective Shields stress is not compatible with the fact to consider a quadratic friction law between the fluid and the sediment interface. Another definition must be set, as the one proposed in \cite{FerMoNarZab}, presented in the introduction by equation (\ref{eq_theta_eff_Q}).

\section*{Acknowledgements}
This research  has been partially supported by the Spanish Government and FEDER through the coordinated Research projects MTM 2015-70490-C2-1-R and MTM 2015-70490-C2-2-R. The authors would like to thank  M.J. Castro D\'{\i}az for the fruitful discussions concerning non-hydrostatic effects.

\end{document}